\documentclass[11pt]{article}

\usepackage{color,helvet,times}
\usepackage{setspace} 
\usepackage{amsmath}
\usepackage{amssymb}
\usepackage{graphicx}
\usepackage{multirow}

\usepackage{natbib} \bibpunct{(}{)}{;}{author-year}{}{,}
\usepackage{amscd}
\usepackage{chemarrow}
\usepackage{verbatim}
\usepackage[normalem]{ulem}

\usepackage{geometry} 
\newcommand{\md}{d\kern-0.035cm\char39\kern-0.08cm}
\newcommand{\mL}{L\kern-0.15cm\char39}

\def\ep{\ensuremath\varepsilon}

\usepackage{calrsfs}

\renewcommand{\subsubsection}[1]{\paragraph{#1.}}

\title{\bf A general approximation for the dynamics of \\quantitative traits}
\author{Katar{\'\i}na  Bo{\md}ov\'a, Ga{\v s}per Tka{\v c}ik, Nicholas H. Barton\\ 
\it  Institute of Science and Technology Austria (IST Austria), \\ \it Am Campus 1, Klosterneuburg A-3400, Austria\\
(Dated: March 12, 2016)
}

\date{}
\begin{document}
\maketitle

\begin{abstract}
Selection, mutation and random drift affect the dynamics of allele
frequencies and consequently of quantitative traits. While the macroscopic dynamics of quantitative
traits can be measured, the underlying allele frequencies are typically unobserved. Can we understand how the macroscopic observables evolve without following these microscopic processes? The problem has previously been studied by analogy with
statistical mechanics: the allele frequency distribution at each time is approximated by the stationary form, which maximises entropy.
We explore the limitations of this method when mutation is small ($4N\!\mu<1$) so that populations are typically close to fixation and we extend the theory in this regime to account for changes in mutation strength. We consider a single diallelic locus under either directional selection, or with over-dominance, and then generalise to multiple unlinked biallelic loci with unequal effects.  We find that the maximum entropy approximation is remarkably accurate, even when mutation and selection change rapidly.

\emph{Keywords:} diffusion approximation, maximum entropy, quantitative genetics

\end{abstract}


Most traits of interest have a complex genetic basis, depending on very many loci.  Quantitative genetics gives a sophisticated statistical description of the components of trait variance, which can predict the immediate change due to selection. The present abundance of genetic markers allows us to find some of the loci that affect traits, but such QTL typically account for only a small fraction of the genetic variance \citep{hill2011,yang2010}.  While we may be able to predict breeding values, and estimate the distribution of effects, it does not seem possible - even in principle - to identify the individual alleles responsible for the bulk of heritable variance.  Thus, we cannot hope to predict the evolution of quantitative traits by using a direct population genetic approach, based on the frequencies of each individual allele.

Here, we develop a general method that allows us to closely approximate the evolution of quantitative traits, knowing only the distribution of allelic effects and mutation rates, but without requiring knowledge of individual allele frequencies.  This can be seen as an extension of the classical infinitesimal model, to include arbitrary gene interactions { and the effects of selection, mutation and drift on the} genetic variance.  It can also be viewed as a  generalized version of the quasi-steady state assumption (QSSA) that is often made in dynamical reaction systems \citep{SS1989,walcher2013} to noisy systems described by PDEs, where the dynamics are approximated using a quasi-stationary distribution assumption (QSDA); here, we use the maximum entropy (ME) principle to define that distribution.

In physics, the maximum entropy principle has a long history, starting with the seminal work of \cite{jaynes1957}, who interpreted the Boltzmann distribution of statistical physics as the most random distribution subject to a constraint on fixed average energy. In the recent decade, there has been a resurgence of interest in ME, especially when applied to biophysics problems ranging from the statistics of neural spiking~\citep{schneidman2006,tkacik2014}, bird flocking~\citep{bialek2012}, protein structure~\citep{weigt2009}, and immunology~\citep{mora2010}. These approaches have been generalized to describe temporal dynamics of high-dimensional systems, known collectively as ``maximum caliber'' or ``dynamical/kinetic maximum entropy'' models~\citep{presse2013}  where the entropy of distribution over temporal paths is maximized given constraints on dynamical variables. Surprisingly, however, ME distributions have not been used widely as a variational ansatz for cases discussed in this work, where the evolution equation for the distribution might be known a priori, e.g., as is the case with the diffusion approximation in population genetics. In such a case the approach differs from the maximum caliber methods since it involves a combination of a static maximum entropy ansatz for the stationary microscopic distribution, with a quasi-stationary assumption in the diffusion/Fokker-Planck equation, which together extend the static maximum entropy inference to a dynamical approximation.

\cite{prugel1994} and  \cite{rattray2001} introduced the maximum entropy approximation to the dynamics of polygenic systems, predicting the cumulants of the trait distribution under mutation, selection and drift; their main motivation was to understand evolutionary algorithms, rather than natural populations.  The same method was described independently in physics \citep{plastino1997}, and used to approximate cosmic ray fluxes \citep{hick1987}.  

However, neither of the two applications of maximum entropy was taken up in their respective fields.  \cite{barton2009} modified Pr\"{u}gel-Bennet, Rattray \& Shapiro's method, so that it could be justified from population genetic principles.  With this modification, it gives the stationary distribution exactly, and is accurate in the limit of slowly changing conditions.  Numerical calculations showed that it is remarkably accurate, even when selection or mutation change abruptly.  However, the method is fully valid only when mutation is stronger than drift ($4N\!\mu>1$), so that populations are almost never fixed for one or other allele.  Yet, in nature  mutation is typically weaker than drift ($4N\!\mu<1$), so that most sites are fixed; in this case, Vladar \& Barton's approximation only applies in cases where mutation or population size do not change through time.

We begin by summarising the stationary maximum entropy approximation and its extension to non-stationary problems. We then extend the maximum entropy approximation so that it applies over the full range of mutation rates, and test the accuracy of this approximation for directional selection and for balancing selection that favours heterozygotes.  This extension is a combination of the continuous approach of \cite{barton2009} with a special treatment of the dynamics at the boundaries. A similar approach, where the boundaries have to be treated differently, has been used in the semi-discrete, semi-continuous methods studied in reaction-diffusion systems \citep{flegg2011,robinson2014}, but also in traveling fitness waves, where the fluctuations can be introduced to the model using a cutoff function (\cite{hallatschek2011,tsimring1996}). We initially consider the distribution of allele frequencies at a single locus, and then extend to multiple loci with a distribution of effects.  Throughout, we assume that recombination is fast relative to other processes, so that the population is in linkage equilibrium, and can be described by its allele frequencies.

\subsection*{Dynamics of allele frequencies} 
The dynamics of allele frequencies $\mathbf p = (p_1,\dots,p_L)$ (for biallelic loci) can be described by a diffusion process using a deterministic forward Kolmogorov  equation { (i.e., the Fokker-Planck or the diffusion equation). The evolution of the joint probability density ${\psi}(\mathbf p,t)$ of allele frequencies satisfies:
	\begin{equation}
		\frac{\partial \psi}{\partial t}  =   -\sum_{i=1}^{ L} \frac{\partial}{\partial p_i}  
		\left[ \mathcal{M}_i(p_i)\psi \right]
		+ \frac{1}{2} \sum_{i=1}^{ L} \frac{\partial^2}{{\partial p_i}^2} \left[ \mathcal{V}_i(p_i) \psi \right] \,, \label{FPE}
	\end{equation}
where the number of loci that contribute to the trait is $L$.

The second term of \eqref{FPE} equals $\mathcal{V}_i(p_i) = \frac{p_i q_i}{2N}$ and captures stochasticity of the allele frequencies arising from random sampling; i.e., the random drift. While in case of linkage disequilibrium this term would contain a double summation, reflecting correlations between loci, the off-diagonal terms vanish at linkage equilibrium.  (The factors of 2 in the brackets of \eqref{FPE} arise because we assume a diploid population of N individuals; the corresponding haploid model would be the same, apart from these factors).

The first term of \eqref{FPE} captures deterministic changes of allele frequencies. { We consider 
\begin{align}
\mathcal{M}_i(p_i) &= (\beta\gamma_i+h \eta_i)p_iq_i \left[1-\frac{2h \eta_i}{\beta\gamma_i+h \eta_i}p_i\right] + \mu q_i - \nu p_i
\end{align}
where $\beta$ is the strength of directional selection, $h$ denotes a higher order correction  that captures dominance, $\mu,\,\nu$ are the forward and backward mutation rates and $\gamma_i$, $\eta_i$ are the additive effects of the $i$-th locus on the traits under selection}. $\mathcal{M}_i(p_i)$ can be written in a potential form 
\begin{align}
\mathcal{M}_i(p_i)& =  \frac{p_i q_i}{2} \frac{\partial (\boldsymbol{ \alpha} \cdot \mathbf{A}) }{\partial p_i}
\end{align} 
where the potential  $\boldsymbol \alpha \cdot \mathbf A$, { obtained by inverting this relationship,} reflects effects of selection and mutation
	\begin{eqnarray}
		\boldsymbol{ \alpha} \cdot \mathbf{A} &=& \sum_{j} \alpha_j A_j 
		= \log \bar W  + \mu U + \nu V \nonumber \\
		&=& \beta \bar z + hH  + 2\mu \sum_{i=1}^{ L}\log p_i + 2\nu \sum_{i=1}^{ L}\log q_i\,,
	\end{eqnarray}
with quantities $\bar z, U, V, H$ are defined as
	\begin{align}
		\bar z &= \sum_{i=1}^{ L}  \gamma_i (p_i  - q_i) \,, \quad H = \sum_{i=1}^{ L} 2\eta_i p_i  q_i \,, \quad
		U = \sum_{i=1}^{ L} 2\log p_i \,, \quad  V = \sum_{i=1}^{ L} 2\log q_i\,.
	\end{align}
where $q_i=1-p_i$ and where $\gamma_i$ and $\eta_i$ are the effects of loci on the traits $\bar z$ and $H$, respectively.

Selection, captured by the mean fitness $\bar W$, may be an arbitrary function of the distribution of quantitative traits, which in turn may be an arbitrary function of the $n$ allele frequencies. The log mean fitness may be further decomposed into a sum in which the $\boldsymbol\alpha_{\bar W}$ represent various selection coefficients and $\mathbf A_{\bar W}$ various selected traits. 
In case of dominance the terms associated to selection are $ \log\bar W =\beta  \bar z  + h  H = \boldsymbol\alpha_{\bar W} \cdot \mathbf A_{\bar W}$ where $\boldsymbol \alpha_{\bar W} = (\beta, h)$ and $\mathbf A_{\bar W} = (\bar z,H)$. Here we assume a weak selection, $\beta \ll1$, such that $\overline{\log W} \approx \log \bar W$. 

The deterministic effects on allele frequency can be summarized   into a vector of coefficients $\boldsymbol \alpha$ and a vector of complementary quantities $\mathbf A$. 
We study directional selection and dominance with non-symmetric mutation, and define $\boldsymbol \alpha$, $\mathbf A$ as
	\begin{equation}
		\boldsymbol \alpha = (\beta, h, \mu, \nu)\,, \qquad \mathbf A = (\bar z, H, U, V)\,. \label{alphaA}
	\end{equation}
In the forthcoming sections, we show that $\mathbf A$ and $\boldsymbol\alpha$ can be understood as constraints and corresponding Lagrange multipliers, respectively, of a particular variational problem.

For $h=0$ and $\mu=\nu$ this represents the simplest scenario of directional selection with symmetric mutation. Directional selection of strength $\beta$ acts on a trait $\bar z$, assumed to be additive while selection of strength $h$ acts on heterozygosity $H$. 
In this work we consider unequal effects $\gamma_i$ on the trait but equal effects $\eta_i=1$ on $H$. This can be easily extended to distribution of effects $\eta_i$. A wide variety of other models can be  treated in the same way. For example, \cite{vladar2011stab} study  stabilising selection on an additive trait.

This diffusion process is known to be an accurate continuous-time approximation to a wide range of specific population genetic models \citep{ewens2004,kimura1955}; moreover, it corresponds directly to the coalescent process that describes the ancestry of samples taken from the population \citep{wakeley2008}.  In order to represent the population in terms of allele frequencies, we must assume that linkage disequilibria are negligible, which will be accurate if recombination is sufficiently fast.  For simplicity, we also assume two alleles per locus. 

The stationary distribution of \eqref{FPE}  has the form  { \citep{wright1931,kimura1955}}
	\begin{eqnarray}
	\bar \psi (\mathbf{p}) &=&  \frac{1}{\mathbb Z} \exp(2N \boldsymbol{ \alpha} \cdot \mathbf{A} ) \prod_{i=1}^{ L} (p_i  q_i)^{-1} 
	=  \frac{1}{\mathbb Z} \bar{W}^{2N} \prod_{i=1}^{ L} p_i ^{4N\!\mu-1}  q_i^{4N\!\nu-1} \label{stationary}
	\end{eqnarray}
where $\mathbb Z$ is a normalization constant, also called a partition function. 
This distribution falls to zero at the boundaries ($p=0,\,1$) provided $4N\!\mu>1$  and $4N\!\nu>1$. However, when mutation rates fall below this threshold, the distribution develops singularities at  boundaries (if  $4N\!\mu$ or $4N\!\nu$ is small the singularity occurs at $p=0$ or $p=1$, respectively), even though the density function is still integrable. 

}

\subsection*{Maximum Entropy in Equilibrium Quantitative Genetics} 

The stationary distribution \eqref{stationary} can be derived from a variational maximum entropy (MaxEnt) principle. The key assumption is that selection and mutation act only through a set of observable quantities, which can be arbitrary functions of allele frequencies, $\mathbf{A} = \{A_i(p)\}$; the strength of selection and mutation are given by the corresponding set of $\boldsymbol \alpha = \{\alpha_i\}$.  Together, these define the potential function $\boldsymbol \alpha \cdot \mathbf A$.

We can define an entropy, relative to a reference measure $\phi(\mathbf{p})$:
	\begin{equation}
		S_H[\psi] = -\int_{[0,1]^{ L}} \psi(\mathbf{p},t)\log\left[ \frac{\psi(\mathbf{p},t)}{\phi(\mathbf{p})}\right] d\mathbf{p}  
		\label{entropy}
	\end{equation} 
which has a unique maximum $\psi(\mathbf{p},t) = \phi(\mathbf{p})$ at the reference distribution; the entropy is (minus) the Kullback-Leibler divergence from $\phi(\mathbf{p})$. The key choice is to set the reference distribution as the neutral distribution of allele frequencies in the absence of mutation or selection:
	\begin{equation}
		\phi(\mathbf{p}) = \prod_{i=1}^{ L} (p_i q_i)^{-1}\,. \label{neutral}
	\end{equation}
Note that $\phi(\mathbf{p})$ is not integrable; however, it does give the neutral probability distribution { given that the allele is not fixed}, and yields the stationary distribution under mutation, selection and drift when we maximise $S_H$ subject to  a normalization constraint and constraints $\langle \mathbf A \rangle = \langle \mathbf A \rangle_\text{obs}$. 
The latter condition enforces a constraint on the ensemble averages $\langle \mathbf A(\mathbf p) \rangle  = \int \mathbf A(\mathbf p) \psi(\mathbf p) d\mathbf p$. { These macroscopic quantities represent information that could in principle be observed. We refer to $\langle \mathbf A(\mathbf p) \rangle$ as observables even though this does not necessarily mean that their values over time are known.}
 The constrained maximization of entropy is solved by a method of Lagrange multipliers; for details, see Appendix~\ref{apLM}. For the example in \eqref{alphaA} the constraints include: 
	\begin{itemize}
		\item[(i)] $ \int_{[0,1]^L}  \psi(\mathbf p) d\mathbf p  = 1$ -- normalization constraint,
		\item[(ii)] $\langle \bar z \rangle =  \langle \bar z \rangle_\text{obs}$ with Lagrange multiplier $2N\!\beta$,
		\item[(iii)] $\langle H \rangle =  \langle H \rangle_\text{obs}$  with Lagrange multiplier $2Nh$,
		\item[{ (iv)}] $\langle U \rangle = U_\text{obs}$ with  Lagrange multiplier $2N\!\mu$,
		\item[{ (v)}] $\langle V \rangle = V_\text{obs}$ with Lagrange multiplier $2N\!\nu$.
	\end{itemize} 
The normalization condition sets the total probability of the allele frequency distribution to one and introduces the partition function $\mathbb Z$ as a constant multiplier in \eqref{stationary}. 

This variational principle recovers the stationary distribution of diffusion equation \eqref{FPE}; the $2N$ times the mutation rates and selection coefficients can be thought of as the Lagrange multipliers. \cite{iwasa1988} introduced the same entropy measure, but used it in a slightly different way: he showed that the sum of the potential function and $S_H/2N$ defines a free fitness that increases through time, just as in thermodynamics the free energy increases in time. Further connections to thermodynamics have been studied by \cite{sella2005} for a special case of very small mutation, where most of the alleles are fixed, and in \cite{bartoncoe2009}, including the novel entropy term (analogous to our $U$, $V$) that involves effects of mutation.

Note that $\mathbf A$ and $\boldsymbol \alpha$ have a specific meaning both in quantitative genetics and in statistical physics. In quantitative genetics,  $\mathbf A$ characterises properties of a quantitative trait whose means can be observed ($\langle \mathbf A \rangle$) and that evolves in response to the evolutionary forces $\boldsymbol \alpha$. In statistical physics, $\mathbf A$ and $\boldsymbol \alpha$ represent conjugate pairs of thermodynamic variables, which can be interpreted as  the constraints and Lagrange multipliers in the variational MaxEnt problem, commonly encountered when microscopic states of the system are unobserved but its macroscopic features are known. 

Note that there is some flexibility in the choice of the reference distribution $\phi(\mathbf p)$, where different choices may lead to the same stationary distribution. For instance, one may take the neutral distribution that involves mutation terms $\phi(\mathbf p) = \prod_k p_i^{4N\!\mu-1}q_i^{4N\!\nu-1}$ while omitting the constraints on $U$ and $V$ and assuming that $\mu, \nu$ are functions of time. This way, the reference distribution would be normalizable. Since these approaches are equivalent, one can view the constraints on $U$ and $V$ as conditions that regularize the allele frequency distribution.

\section*{Dynamic Maximum Entropy Approximation}

Our aim is to approximate the dynamics of a high-dimensional system by a small number of variables, which include those quantities that determine fitness. We approximate the real distribution of allele frequencies by the stationary distribution obtained by the  MaxEnt method with a small number of constraints and use it as an ansatz in the diffusion equation. This leads to effective dynamical forces $\boldsymbol \alpha^\ast$ that yield the correct dynamics for the observed quantities. The assumption that the population is perturbed only through the forces $\boldsymbol \alpha$ is crucial to the success of our approximation.  If we could manipulate individual allele frequencies in an arbitrary way, then the long-term evolution would become essentially unpredictable: initially rare alleles could increase to cause arbitrary changes as they eventually rose to appreciable frequency (\cite{barton2009}, Fig. 1). The overall strategy of the dynamical MaxEnt (DynMaxEnt) approximation is summarized in Table~\ref{tab2} while the terminology from  statistical physics and quantitative genetics is in Table~\ref{tab_terminology}. Various approximate methods,  discussed in our work, are summarized in Table~\ref{tab3}.
	\begin{table}[t]
	\begin{center}
	\caption{Summary of the DynMaxEnt approach in four steps.\label{tab2}}
	\begin{tabular}{|c|p{10cm}|}
		\hline
		{\bf  Step 1} & 
		{ Formulate dynamics, as in \eqref{FPE}, for the probability distribution of the state variables 
		$\psi(p)$}.\\\hline
		{\bf  Step 2} &
		{  Obtain the stationary distribution $\psi$ and write it in an exponential (log-linear) form 
		$\psi(p) \propto \phi (p)\exp{(2N \boldsymbol \alpha \cdot \mathbf A)}$ in terms of observables
		$\langle \mathbf A \rangle$ and constant forces $\boldsymbol \alpha$.}\\\hline
		{\bf  Step 3} & 
		{ Represent $\psi(p)$ as a solution of a variational MaxEnt problem with reference distribution $\phi(p)$, 
		constraints on  $\langle \mathbf A \rangle$ and Lagrange multipliers $\boldsymbol \alpha$ (non-unique). }\\\hline
		{\bf  Step 4} & 
		{ Use a quasi-stationarity assumption to approximate dynamics of observables using the stationary 
		distribution  where the coefficients $\boldsymbol \alpha$ are alowed to change in time to match the correct 
		dynamics of observables. This criterion leads to a reduced dynamical system for the effective coefficients 
		$\boldsymbol \alpha^\ast$.}
		\\\hline
	\end{tabular}
	\end{center}
	\end{table}

{We first describe the continuous DynMaxEnt method, proposed in \cite{barton2009}, that requires a sufficient number of mutations in every generation.  On the other hand, a discrete approximation, also used in \cite{barton2009} and described in Appendix~\ref{apD}, is applicable when the mutation rate is small and selection is limited.} The  dynamics are then
formulated in terms of  fixed classes of alleles. However, the discrete approximation is not accurate unless mutation { rate} is very small and even then it has a limitation when $N\!\beta \gg1$ (Appendix~\ref{apD}). Similarly, the continuous method fails for $4N\!\mu <1$ (Appendix~\ref{apC}). We compare the performance of these methods in Appendix~\ref{apComp} and find that while the discrete method applies to very small mutation { rate} and the continuous method to large mutation { rate}, the intermediate regime is not captured by either of them.  This  is similar to the result of \cite{mustonen2007,mustonen2008} who studied fitness waves in the problem of fluctuating selection and introduced a novel approximation for a problem of fluctuating selection, accurate for small selection timescale. Figure 2  in \cite{mustonen2008} also shows an intermediate regime where neither the diffusion theory, nor the novel approximation are accurate.  In this work we present a \emph{general dynamical MaxEnt} approximation that is applicable in all regimes. 
{This aproximation is compared to the numerical solution of the diffusion equation. Instead of using individual based simulations, which are computationally demanding, we consider allele frequencies to hold biologically relevant values, $p_k = k/2N$ for $k=0,\dots, N$, and forward iterate an explicit transition matrix, consistent with the diffusion equation. This approach is feasible also when a moderate number of loci have different effect on the trait of interest.}
	\begin{table}[t]
	\begin{center}
	\caption{Table of key terms and constants.\label{tab_terminology}}
	\begin{tabular}{|c|p{10cm}|}
		\hline
		Symbol & Quantitative genetics / Statistical physics \\\hline \hline
		$ \mathbf A $ & Macroscopic observables \\\hline
		$\langle \mathbf A \rangle$ &  { Constraints}  \\\hline
		$\boldsymbol \alpha$ & Evolutionary forces  / Lagrange multipliers \\\hline
		$\psi(p,t)$ & Distribution of allele frequencies / { Boltzmann} distribution \\\hline
		$\phi(p)$ & Neutral distribution (no mutation)  / Reference distribution \\\hline
		$S_H$ &  Relative entropy, negative Kullback-Leibler divergence \\\hline
		$\bar W$ & Mean fitness    \\\hline
		$2N$ & Population size in a diploid population   \\\hline
		${ L}$ & Number of loci contributing to the trait   \\\hline
		$B$ & Additive genetic covariance matrix  \\\hline
		$C$ & Covariance in fluctuations / Susceptibility matrix  \\\hline
		$J[\psi,p]$ & Flux of the probability mass at frequency $p$   \\\hline
	\end{tabular}
	\end{center}
	\end{table}

	\begin{table}[t]
	\begin{center}
	\caption{ Overview of the approximations. \label{tab3}}
	\begin{tabular}{|p{0.25\textwidth}|p{0.7\textwidth}|}
		\hline
		{\bf  Abbreviation} & 
		{\bf  General Approach}\\  \hline
		{ MaxEnt} & { Stationary  microscopic distribution from macroscopic observables.} \\\hline
		{ DynMaxEnt} & {  Any approximation for the dynamics of observables based on a combination of MaxEnt variational ansatz, quasi-stationarity assumption, and a dynamical (e.g. diffusion) equation for the probability distribution. }\\\hline
		{\bf  Abbreviation} & 
		{\bf  Method in Quantitative Genetics} \\  \hline
		{ Continuous DynMaxEnt} & {Refers to the method of \cite{barton2009} where the dynamics are captured by diffusion 
		equation \eqref{FPE}. } \\\hline
		{ General DynMaxEnt} & The approximation introduced here,  which generalizes the Continuous DynMaxEnt method. This new approximation distinguishes between the bulk mass that behaves as 
		in the continuous problem, and the coupled boundary masses, that behave as discrete quantities. \\\hline
	\end{tabular}
	\end{center}
	\end{table}
\subsection*{ Continuous DynMaxEnt approximation} \label{secC}

Any set of forces, $\boldsymbol \alpha$, will cause the population to evolve to a stationary distribution $\frac{\phi}{\mathbb Z} \exp(2N \boldsymbol\alpha\!\cdot\!\mathbf A)$; this is the distribution that maximises entropy subject to constraints on $\langle \mathbf A \rangle$, the $2N\boldsymbol \alpha$ being the Lagrange multipliers.  Now, suppose that the forces change abruptly, from $\boldsymbol \alpha_0$ to $\boldsymbol \alpha_1$ and no further information about the system is provided. The expected observables, $\langle \mathbf A \rangle$, will evolve towards the new stationary distribution.  At any time, there will be a set of forces $\boldsymbol \alpha^\ast$ that would produce the current $\langle \mathbf A \rangle$ if the population were stationary; we expect that the $\boldsymbol \alpha^\ast$ will evolve from $\boldsymbol \alpha_0$ to $\boldsymbol \alpha_1$, as the population evolves from one stationary state to the other.  Thus, we can describe the dynamics either by the change in $\langle \mathbf A \rangle$, or equivalently, by the change in the $\boldsymbol \alpha^\ast$.

Under the diffusion approximation, the expectations change as:
	\begin{align}
		\frac{\partial \langle A_i \rangle}{\partial t} &= \sum_j B_{i,j} \alpha_j + \frac{1}{2N} V_i
	  \label{expectations}
	\end{align}
where
	\begin{align}
		\quad B_{i,j} &= \left \langle \sum_k \frac{\partial A_i}{\partial p_k} \frac{p_k q_k}{2} \frac{\partial A_j}{\partial p_k} 
		\right \rangle\,, \quad V_i = \left \langle \sum_k \frac{\partial^2 \!A_i}{\partial p_k^2} \frac{p_k q_k}{2}  \right \rangle \label{matrixB}
	\end{align}
(Eqs. 13, 14 of  \cite{barton2009}; note that in their Eq. 13, the expectation should be taken over the whole equation, not inside the derivatives as typed).  The expectations that appear on the left-hand side of \eqref{expectations} are not the same as the ones on the right, therefore the system is not closed. We now introduce the continuous dynamic maximum entropy approximation (cont. DynMaxEnt), namely, that the $B_{i,j}$, $V_i$ in the dynamically changing system are approximated by the values that they would have at the corresponding stationary state that generates the actual $\langle \mathbf A \rangle$; the stationary distribution coincides with the MaxEnt distribution. If the population were at a stationary state under the forces $\boldsymbol \alpha^\ast$, chosen to produce the current expectations $\langle \mathbf A \rangle$, then there would be no change:
	\begin{align}
		\frac{\partial \langle A_i \rangle}{\partial t} &= 0 = \sum_j B_{i,j}^\ast \alpha_j^\ast + \frac{1}{2N} V_i^\ast \label{QS}
	\end{align}
where the $^\ast$ denotes values at the stationary distribution. The approximation has a form $ \frac{1}{2N} V_i^\ast \approx - \sum_j B_{i,j}^\ast \alpha_j^\ast$ which gives  Eqn. 15 of  \cite{barton2009}:
	\begin{equation}\label{BE}
		\frac{\partial \langle A_i \rangle}{\partial t} = \sum_j B_{i,j}^{ \ast} (\alpha_j - \alpha_j^\ast)\,.
	\end{equation}
It may be more convenient to follow the rates of change in the $\boldsymbol \alpha^\ast$, which can be written in terms of the covariance of fluctuations in the $\mathbf A$. Using matrix notation:
	\begin{eqnarray} \label{QSDA}
		\frac{\partial \boldsymbol {\alpha^\ast}}{\partial t} &=&  \frac{1}{2N} { \mathbf C}^{\ast-1} {\mathbf B}^\ast \cdot 
		(\boldsymbol{ \alpha} - \boldsymbol {\alpha^\ast}) 
	\end{eqnarray}
where
	\begin{eqnarray}
			{ \mathbf C}^\ast& =& \text{Cov}(A_i,A_j) = \left\langle \frac{\partial^2 \log \mathbb Z}{\partial \alpha_i \partial \alpha_j} \right \rangle \,, \label{matrixC}
	\end{eqnarray}
with an initial condition $\boldsymbol \alpha^\ast(0) = \boldsymbol\alpha_0$ and $\boldsymbol \alpha = \boldsymbol\alpha_1$. The difference $|\boldsymbol \alpha_1-\boldsymbol \alpha_0|$ represents the change in evolutionary forces.
Since the matrices $\mathbf B^\ast$ and $\mathbf C^\ast$ depend only on the effective forces $\boldsymbol \alpha^\ast$, as shown in Appendix~\ref{apBC}, the dynamical system for $\boldsymbol \alpha^\ast$ is closed. A detailed derivation of the DynMaxEnt method under more general conditions can be found in Appendix~\ref{apDerivation}.

Intuitively, one may assume quasi-stationarity in \eqref{expectations}, provided that the evolutionary forces $\boldsymbol\alpha$ change on a slower timescale than the time scales of selection ($1/\beta$), mutation ($1/\mu$) and random drift ($2N$). Then, the adiabatic approximation in \eqref{QSDA} should be accurate not only for the predicted observables but also for the microscopic distribution. However, we will show that even when the  evolutionary forces change abruptly, i.e. when $|\boldsymbol \alpha_1-\boldsymbol \alpha_0|$ is not small, the approximation remains accurate -- even though there is no guarantee that the inferred microscopic distribution agrees with the correct one. 

The matrix $\mathbf B$ can be seen as a generalization of the additive genetic covariance matrix, where the $\partial A_i/\partial p_k$ correspond to (twice) the marginal effects of the k'th allele on the $A_i$.  The maximum entropy approximation consists in assuming that this matrix is approximately what one would obtain at equilibrium with the current $\langle \mathbf A \rangle$. Thus, \eqref{BE}  is an extension to the "breeder's equation"  \citep{walsh1998} which allows for quantities $\langle \mathbf A \rangle$ that can be any function of allele frequencies -- not just trait means -- and that allows for random fluctuations, mutational bias and non-additive selection. 

The DynMaxEnt method can be contrasted with the maximum caliber method, reviewed in \cite{presse2013}. DynMaxEnt uses static observables to infer the correct stationary allele frequency distribution, but allows the Lagrange multipliers to change in time in accordance with the known diffusion equation ensuring that the observables  are correct at all times. On the other hand, the maximum caliber method uses constraints on temporal characteristics to arrive at a distribution over the allele frequency paths with constant values of Lagrange multipliers. DynMaxEnt is suitable for our problem since it only assumes knowledge of initial and changed evolutionary parameters and no further information on the properties of the allele frequency paths.

\subsection*{General DynMaxEnt approximation} \label{secG}

When numbers of mutations are small (i.e. $4N\!\mu<1$, $4N\!\nu<1$),  we  face a problem of diverging components in the {continuous} DynMaxEnt approximation due to a U-shaped  allele frequency distribution (for simplicity, we will consider a single locus). If $\mathbf A = (\bar z, H,U,V)$ this divergence is caused by diagonal elements of matrix $\mathbf B$ that correspond to $U$ and $V$, in particular $B_{ 3,3} = \langle pq/2 \cdot (\partial_{p} U(p))^2 \rangle= \langle 2q/p \rangle$ for $4N\!\mu<1$ and $B_{4,4} = \langle pq/2 \cdot (\partial_{p} V(p))^2 \rangle = \langle 2p/q \rangle$  for $4N\!\nu < 1$ (see Appendix~\ref{apBC}, \ref{apC} for more detail). 
Therefore the continuous DynMaxEnt approximation fails completely in a regime of dynamic selection and mutation when $4N\!\mu<1$ or $4N\!\nu<1$ simply because the right hand side of the dynamical system \eqref{QSDA} is ill-defined. {The breakdown of the continuous DynMaxEnt method, when number of mutations are small (i.e. for small populations), is not a numerical problem but a fundamental limitation of the method itself.} 
However, it can be avoided by considering a modified diffusion problem which does not aim to resolve all details of the allele frequency distribution close to the fixation and loss, but instead agrees with the original problem in terms of the probability that the allele frequency is extreme. 

We define the boundary layers as $[0,\ep)$ and $(1-\ep,1]$ for $\ep\ll 1$ but finite. {The value of the truncation parameter $\ep$ is discussed later but typically $\ep \sim 1/N$.} We then replace the original diffusion equation with solution $\psi(p,t)$  by a new system of partial differential equations with solution $\psi_\ep(p,t)$ that agrees with the true dynamics in the following properties:
	\begin{itemize}
		\item[(a)] The stationary distribution in the bulk is the same for both problems: 
		$\psi_\ep(p,t=\infty) = \psi(p,t=\infty)$ for $ p\in[\ep,1-\ep]$.

		\item[(b)] The stationary probabilities of extreme allele frequencies are the same for both problems:
		$$P[p<\ep] = \int_0^\ep \psi(p,t=\infty) = \int_0^\ep \psi_\ep(p,t=\infty)$$ 
		and also 
		$$P[p>1-\ep] = \int_{1-\ep}^1 \psi(p,t=\infty) = \int_{1-\ep}^1 \psi_\ep(p,t=\infty)$$  
		to the {lowest} order in $\ep$. 

		\item[(c)] As the truncation parameter goes to 0, the problem converges to the original diffusion equation, 
		i.e. it develops singularities at the boundaries:
		$$\lim_{\ep\rightarrow 0} \psi_\ep(p,t) = \psi(p,t)\,.$$
	\end{itemize}
Replacing the original diffusion equation by a set of coupled diffusion equations in different regions of the state space { captures an important characteristic of the problem: the presence of multiple timescales in the allele frequency dynamics.} When the system is perturbed from stationarity by a change in the Lagrange multipliers (selection, mutation, heterozygosity), for instance by a dramatic drop in mutation, the correct distribution very quickly 
{ adjusts to the form $p^{4N\!\mu-1}q^{4N\!\nu-1}$ at the boundaries. Only then does the mass in the interior slowly transfer} 
to the vicinity of the fixed states to converge to the new stationary distribution. 
This results in a very fast dynamics of $\langle U\rangle$ and $\langle V\rangle$, and a considerably slower dynamics of the trait mean and the mean heterozygosity. 

DynMaxEnt can capture these multiscale features only if it can incorporate a
{ low and changing rate of mutation}. 
The quick initial adjustment of the mutation rates is then naturally followed by a slow dynamics of the trait mean and heterozygosity, because the speed of the transfer of the mass between the fixed states is limited by 
{ the infrequent rate of mutation}. 
{ Similarly to other observables, the boundary masses can also be expressed as means of functions of allele frequency and thus treated as additional observables. The appropriate function is a characteristic function $\chi_{[a,b]}$, where
	\begin{equation}\label{constraints}
		\chi_{[a,b]} = 
		\begin{cases}
			1 & p\in [a,b]\,,\\
			0 & \text{else}\,.
		\end{cases} 
	\end{equation}
}
Splitting the problem {domain} augments the  degrees of freedom by  $\langle \chi_{[0,\ep]}\rangle$ and $\langle \chi_{[1-\ep,1]} \rangle$, which control the boundary dynamics independently of the bulk dynamics,
and the corresponding Lagrange multipliers $2N\!\kappa$ and $2N\!\rho$.  The values of $\kappa$ and $\rho$ will be determined later.
The general DynMaxEnt method, derived in Appendix~\ref{apDerivation} and summarised below, provides a way to couple the boundary dynamics with the bulk dynamics to account for the multiscale features,  and resolve the technical problems of the continuous method while keeping the same number of degrees of freedom as the continuous DynMaxEnt.

We first split the allele frequency domain into the bulk part and the boundary part
and define the diffusion equation separately in the three regions in Appendix~\ref{apDerivation}. We couple the equations by a boundary flux that is  consistent with the original diffusion equation, thus leading to the same probability mass at the boundaries as in the continuous DynMaxEnt method. The stationary distribution of the problem \eqref{PDE2} has the form
	\begin{equation} \label{SCdist2}
		\bar \psi_\ep( p) =   \frac{1 }{\mathbb Z}  \bar W^{2N}
		\begin{cases}
			e^{2N\!\kappa}\ep^{4N\!\mu-1}  q^{4N\!\nu-1} & \text{if $p<\varepsilon$}\\
			\phantom{e^{2N\!\kappa}}  p^{4N\!\mu-1} q^{4N\!\nu-1} & \text{otherwise}\\
			e^{2N\!\rho}p^{4N\!\mu-1} \ep^{4N\!\nu-1} & \text{if $p>1-\varepsilon$}
		\end{cases}\,,
	\end{equation}
with the normalization constant $\mathbb Z$ such that $1 =  \int \psi_\ep( p)d p$ and the relative masses in the three regions are determined by constants $\kappa$ and $\rho$.
The stationary distribution \eqref{SCdist2} is not generally continuous at $p=\ep,1-\ep$. However, it can still be obtained by maximising a relative entropy with a bounded base distribution
	\begin{equation}\label{neutral}
		\phi_\ep(p)= 
		\frac{1}{\mathbb Z^0}
		\begin{cases}
			(\ep q)^{-1} & \text{if $p<\varepsilon$}\\
			(p q)^{-1} & \text{otherwise}\\
			(p \ep)^{-1} & \text{if $p>1-\varepsilon$}
		\end{cases}\,,
		\quad \mathbb Z^0 =  \int_0^{1} \phi_\ep (p)dp\,,
	\end{equation}
complemented by the following constraints on $\langle \mathbf A \rangle$ and Lagrange multipliers $2N\!\boldsymbol \alpha$
	\begin{align}
		\mathbf A &= ( \log\bar W, U_\ep , V_{\ep} ) \,, \label{A}\\
		\boldsymbol \alpha &= \left(\!\boldsymbol\alpha_{\bar W} ,\!\mu,\!\nu \right)\,, \label{alpha}
	\end{align}
where instead of $U$ and $V$ that diverge at the boundaries we take their truncation to $p\in[\ep,1-\ep]$
	\begin{align}\label{constraints}
		U_\ep = 
		\begin{cases}
			2\log \ep & p<\ep\\
			2\log p & \text{else}
		\end{cases} \,, \quad
		V_{\ep} = 
		\begin{cases}
			2\log q & p<1-\ep\\
			2\log \ep & \text{else}
		\end{cases} \,.
	\end{align}
The two remaining parameters $\kappa$ and $\rho$ are matched to satisfy conditions (b), leading to 
	\begin{align}\label{SCdist4}
		\kappa = -\frac{1}{2N} \log (4N\!\mu) \qquad \text{and} \qquad \rho = -\frac{1}{2N} \log (4N\!\nu)\,.
	\end{align}
This relationship ensures that the approximate stationary distribution has the same proportion of the mass at the boundaries to the stationary solution of the original diffusion. 
Generalized to multiple loci, the MaxEnt distribution has the form
	\begin{align} \label{SCdist}
		\bar \psi_\ep (\mathbf p) =   \frac{1 }{\mathbb Z} \bar W^{2N} \prod_{i=1}^{ L}
		\begin{cases}
			\frac{1}{4N\! \mu } \ep^{4N\!\mu-1} q_i^{4N\!\nu-1} 
			& \text{if $p_i<\varepsilon$}\\
			\phantom{\frac{1}{4N\! \mu } }p_i^{4N\!\mu-1} q_i^{4N\!\nu-1} & \text{otherwise}\\
			\frac{1}{4N\!\nu} p_i^{4N\!\mu-1} \ep^{4N\!\nu-1} & \text{if $p_i>1-\varepsilon$}
		\end{cases}\,,
	\end{align}
with $\mathbb Z$ such that $1 =  \int \bar \psi_\ep (\mathbf p)d\mathbf p$.
We remark that the reference distribution $\phi_\ep(p)$ is the stationary distribution in the absence of selection and mutation. Therefore, $\phi_\ep(p) = \sigma_\ep(p)^{-1}$ where $\sigma_\ep$ is the state-dependent diffusion coefficient
	\begin{align}
		\sigma_\ep(p)& = 
		\begin{cases}
			\ep q & p<\ep\\
			pq & \text{else}\\
			p\ep & p>1-\ep
		\end{cases} \,.
	\end{align}

The diffusion equation in the split domain can be used to derive a new DynMaxEnt approximation for arbitrary mutation strengths. This is done similarly as in the continuous DynMaxEnt: using the ansatz \eqref{SCdist2} and a quasi-stationarity assumption. The derivation, detailed in Appendix~\ref{apDerivation}, leads to the same dynamics as the original method
	\begin{align}
		\frac{\partial \langle \mathbf {A} \rangle}{\partial t}& = {\mathbf B}_\ep (\boldsymbol {\alpha} - \boldsymbol {\alpha}^\ast) 
		\label{eqAep}\,, \\
		\frac{\partial \boldsymbol { \alpha}^\ast}{\partial t}&= \frac{1}{2N}   
		{\mathbf C}_\ep^{-1} {\mathbf B}_\ep( \boldsymbol {\alpha}-\boldsymbol { \alpha}^\ast)\,
		\label{eqaep}
	\end{align} 
where \eqref{eqaep} is a closed dynamical system for $\boldsymbol \alpha^\ast$.
However, since constant prefactors in the stationary distribution \eqref{SCdist} depend on mutation rates, the matrix ${\mathbf C}_\ep$ contains additional terms
	\begin{align}\label{BCeps}
		{\mathbf B}_{\ep} &= \left \langle \frac{{ \sigma_\ep}(p)}{2} \frac{\partial A_j}{\partial p} \frac{\partial A_k}{\partial p} \right \rangle
		\,,\nonumber \\
		{\mathbf C}_{\ep} &= \text{Cov}(A_j,A_k) - \nonumber \\
		&\begin{pmatrix}
			\mathbf 0_{\bar W} &  \frac{1}{2N\!\mu} \text{Cov}(\mathbf A,\chi_{[0,\ep)}) 
			&\frac{1}{2N\!\nu} \text{Cov}(\mathbf A,\chi_{[1-\ep,1)}) 
		\end{pmatrix},
	\end{align}
where the subtracted terms form a square matrix of the same dimension as $\text{Cov}(A_j,A_k)$. Its leftmost columns, denoted by $\mathbf 0_{\bar W}$, represent a contribution to the terms $\text{Cov}(\cdot ,\mathbf A_{\bar W})$. Since the stationary distribution \eqref{SCdist} depends on the selection-related Lagrange multipliers only through $\bar W^{2N} = e^{\boldsymbol \alpha_{\bar W} \cdot \mathbf A_{\bar W}}$, similarly to the continuous DynMaxEnt method, all added terms are zero. In contrast, presence of  the mutation-related Lagrange multipliers in \eqref{SCdist} as the scaling factors results in additional terms added to $\text{Cov}(\cdot ,U_\ep)$ and $\text{Cov}(\cdot ,V_\ep)$, forming the remaining two columns in \eqref{BCeps}. The components of the matrices $\mathbf B_\ep$ and $\mathbf C_\ep$ are functions of effective forces only, similar to the continuous DynMaxEnt approximation. The difference is that the expectations involve integrals through parts of the domain $[0,1]$ (bulk part or boundary parts) and thus the terms have to be evaluated numerically.

\section*{Applications of the general DynMaxEnt method}

In the following, we show the performance of the general DynMaxEnt method in the parameter regimes when the continuous DynMaxEnt is not applicable, because the dynamics \eqref{QSDA} become singular. These examples involve dynamic selection and mutation in the regime $4N\!\mu, 4N\!\nu < 1$. 

Performance of the DynMaxEnt approximation for different scenarios is compared to the discrete and continuous approximations in Appendix~\ref{apComp}.

\subsection*{Example 1: { Low and changing} mutation}

Here we consider a single locus under {directional selection. As we show in Appendix E, this also corresponds to the case of multiple loci with equal effects.} 
Despite its apparent simplicity, the system still contains $2N$ degrees of freedom that capture the allele frequency distribution in a population of $N$ diploid individuals. The general DynMaxEnt approximation, based on a stationary MaxEnt approximation, reduces  the dynamics to a few degrees of freedom that correspond to the Lagrange multipliers $\boldsymbol \alpha^\ast(t)$. For instance when $\mathbf A = (\bar z, U_\ep,V_\ep)$ the full dynamics in the $2N$-dimensional space reduces to  a 3-dimensional space $(\beta^\ast,\mu^\ast,\nu^\ast)$.

The general DynMaxEnt method is tested in the most challenging situation when the initially strong mutation suddenly changes to $4N\!\mu \ll1$. 
The continuous DynMaxEnt method does not apply to this example due to two reasons. First, when effective mutation drops below $4N\!\mu^\ast=1$ or $4N\!\nu^\ast = 1$, the components of matrix $B$ diverge due to singularities of the stationary allele frequency distribution.
Second, even when the mutation rates are kept fixed at their terminal values ($\mu^\ast = \mu$, $\nu^\ast=\nu$), resulting in a reduced dynamics, the continuous DynMaxEnt does not give correct dynamical predictions for small mutation rates.  A closer look at the failures of the continuous DynMaxEnt method is in Appendix~\ref{apC}.

\begin{figure}[h]
		\centering
		\includegraphics[width=0.97\textwidth,keepaspectratio]{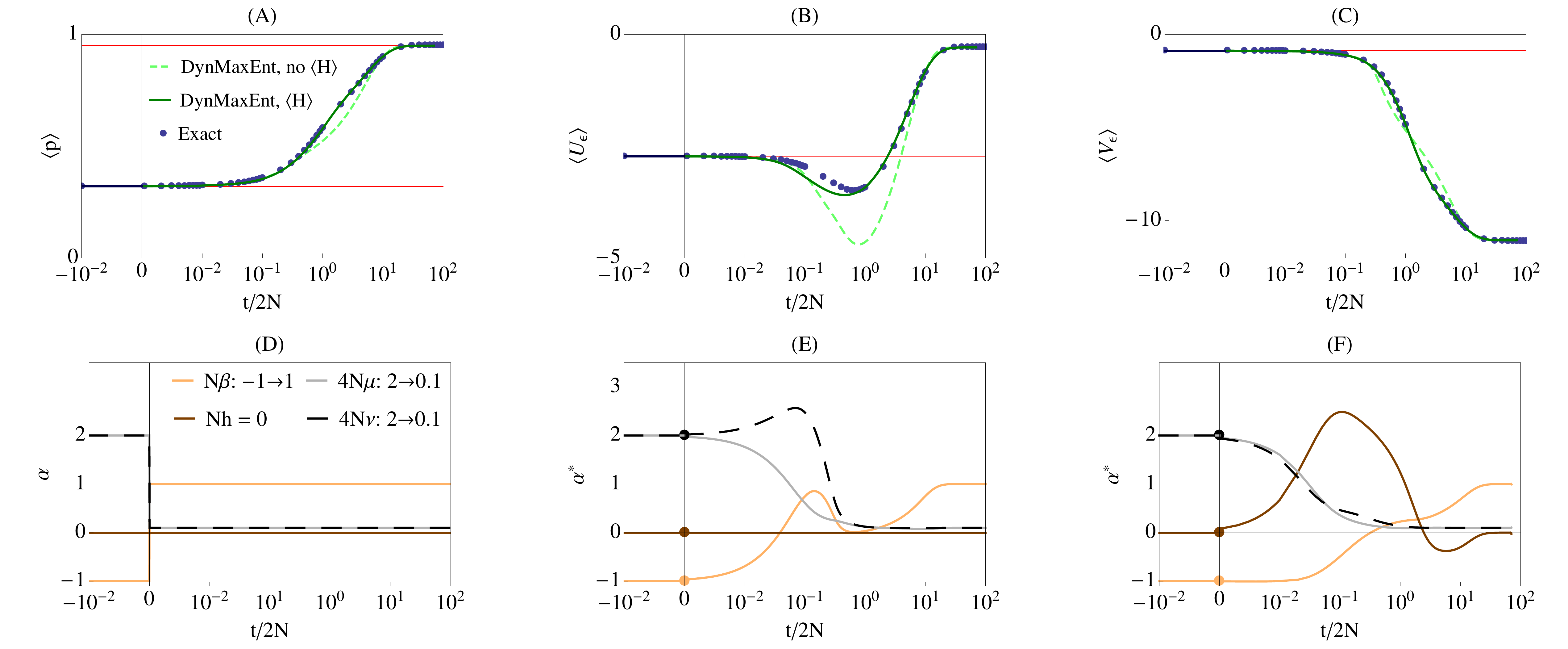}
		\caption{  Example 1, details of the general DynMaxEnt method. 
 			The response of the observed quantities: panels (A), (B) and (C) display observables $\langle p \rangle$,  
			$\langle U_\ep \rangle$ and 
			$\langle V_\ep \rangle$, obtained by numerically solving the diffusion equation using a discretization of space 
			and time where the explicit transition matrix is known, in blue dots and the general DynMaxEnt approximations in green color. 
			We used $2N=1000$ and $\ep = 1/2N$.
			The dashed green shows approximation with the three shown observables and the solid green uses 
			an additional observable  $\langle pq \rangle$.
			(D) Changes in evolutionary forces that draw the system out of equilibrium 
			include a rapid decrease in mutation, complemented 
			by a change in selection strength. 
			(E) Dynamics of Lagrange multipliers in the general DynMaxEnt method when three constraints are employed: 
			$\langle p \rangle$,   $\langle U_\ep \rangle$ and 
			$\langle V_\ep \rangle$. 
			(F) Dynamics of Lagrange multipliers in the general DynMaxEnt method when four constraints are employed, including $\langle pq \rangle$.
		\label{fig_ex1f3} }
	\end{figure}

The general method (with $\ep=1/2N$) gives a satisfactory estimate of the trait mean at the beginning of the adaptation process, shown in Figure~\ref{fig_ex1f3}, while a quick drop in effective mutation, compensated by a change in the effective selection, results in a quick loss of polymorphism. When the equilibration of the effective mutation is too fast, the speed of the dynamics which are limited by effective mutation becomes slower than necessary and leads to underestimation of trait mean and $\langle U_\ep \rangle$.
This can be observed over the medium timescale $\frac{t}{2N} \sim 1-10$.

Over long timescales, the approximate dynamics accurately match the exact dynamics. When we employ an additional constraint on the mean heterozygosity, although no a priori selection is acting on it, we increase the number of degrees of freedom from three to four. While a general feature of variational approximations is that increase in the number of constraints leads to an improvement in the performance of the approximation,  this particular choice yields a match that is almost indistinguishable in all four observables (including the added $\langle pq \rangle$), as shown in Figure~\ref{fig_ex1f3}. 

The dynamics of Lagrange multipliers in Figure~\ref{fig_ex1f3} (E-F) demonstrates separation of timescales. An initial quick drop in mutation  $\nu^\ast$ and subsequently $\mu^\ast$ to the vicinity of their steady states is compensated by the changes in  $\beta^\ast$ (and $h^\ast$) and followed by a slow relaxation of $\beta^\ast$ (and $h^\ast$) to equilibrium once the mutation rates are barely changing. It is interesting that the dynamics of the effective forces differs significantly between cases (E) and (F).

\subsection*{Example 2: Overdominance}

Here we consider that the population dynamics is driven not only by directional selection but also by selection on heterozygosity, i.e. overdominance. This yields stationary distributions that may have up to three modes: two modes represent peaks at the boundaries in the low mutation regime and an intermediate peak represents increased probability of heterozygous individuals. We consider an analogous example to Example 1 but add initially strong selection on heterozygosity (trimodal distribution of allele frequency) that switches to a purely directional selection. 

	\begin{figure}[h]
		\centering
		\includegraphics[width=0.97\textwidth,keepaspectratio]{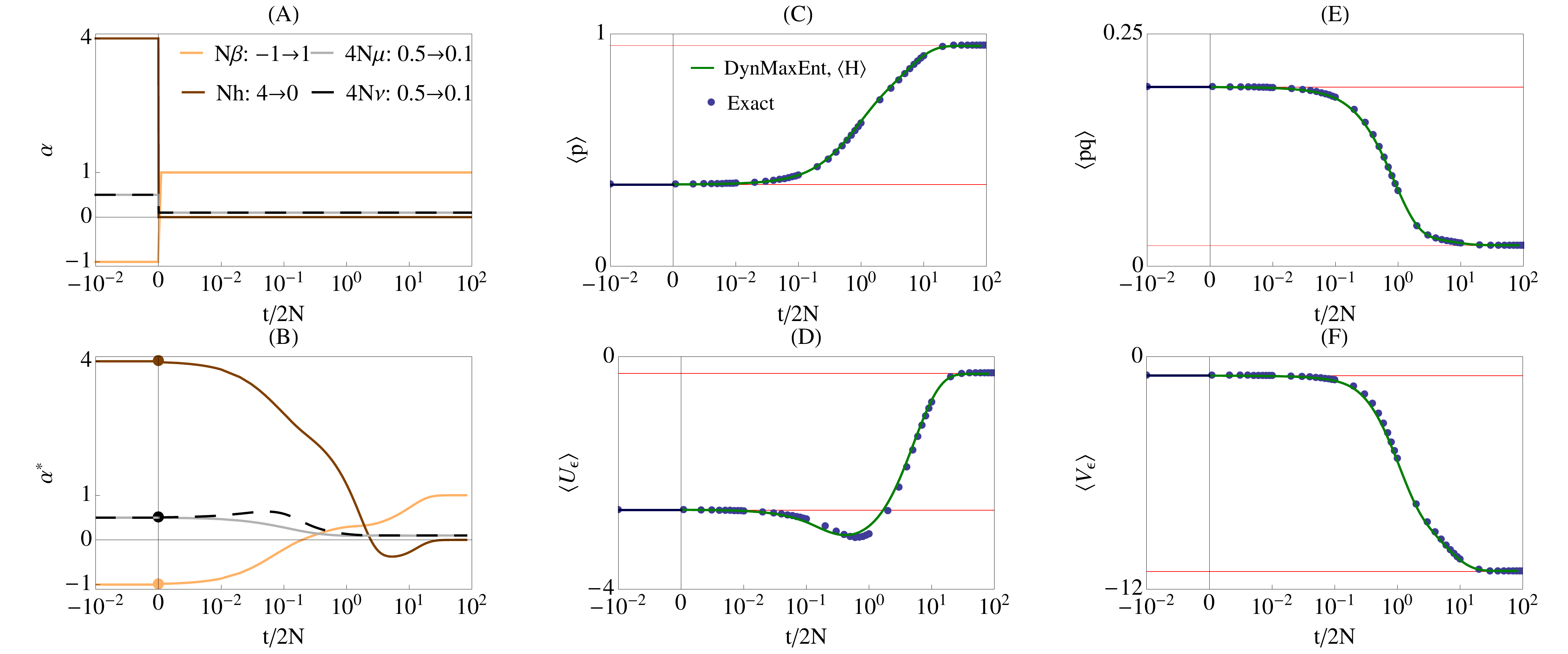}
		\caption{  {Example 2, details of the general DynMaxEnt method: 
			(A) Changes in evolutionary forces, including a rapid decrease in mutation, complemented 
			by a change in selection strength and a change in the selection strength on heterozygosity. 
			(B) The effective forces $\boldsymbol \alpha^\ast$. (C-F)
			The response of the observed quantities 
			$\langle p \rangle$, $\langle pq \rangle$,  { $\langle U_\ep \rangle$} and 
			{$\langle V_\ep \rangle$}.  Exact dynamics, obtained by numerically solving the diffusion equation 
			using a discretization of space and time where the explicit transition matrix is known, are
			shown in blue dots, the  general DynMaxEnt} approximation is shown in green. 
			We used $2N=1000$ and $\ep = 1/2N$.
		\label{fig_ex2f1} }
	\end{figure}

Figure~\ref{fig_ex2f1} shows an almost perfect agreement between the exact solution and the approximation despite the fact that the approximation is not built to capture rapid switches in the forces but, { to the} contrary, assumes that the forces change slowly. The effective forces do not capture the rapid change in the real forces on the system, but  show a slow-fast relaxation to their new steady states. Appendix~\ref{apI} shows a complementary simulation when the initial ($\boldsymbol \alpha_0$) and terminal conditions ($\boldsymbol \alpha_1$) are switched. The system follows a different trajectory in the direction $\boldsymbol \alpha_0 \rightarrow \boldsymbol \alpha_1$ and $\boldsymbol \alpha_1 \rightarrow \boldsymbol \alpha_0$.  

In all simulations, presented here, we use $2N\!\ep = 1$; such a threshold $\ep$ corresponds to an allele frequency of a single individual out of a population of size $2N$.  In Appendix~\ref{apep}, we provide additional simulations to show that there exists an optimal threshold $N\ep$, which minimizes the approximation error. Moreover, we show that in the worst case scenario, where a strong change in selection is coincident with a low and changing mutation, the relative error of the approximation is on the order of 1 percent.

\subsection*{Example 3: Multiple loci with different effects}

In the following example, we show  the  applicability of the DynMaxEnt approximation for the evolution of  quantitative traits that depend on multiple loci with different effects. The state space of the system of $L$ independent loci contains essentially $(2N)^L$ degrees of freedom since each locus is characterized by allele frequency distribution with $2N$ degrees of freedom. The reduction of the dimensionality to three (directional selection) or four (overdominance) degrees of freedom thus offers an immense simplification of the problem  where instead of tracking the full allele frequency distribution we are tracking just the dynamics of the Lagrange multipliers $\boldsymbol\alpha^\ast$, corresponding to the underlying constraints. 

How does the method perform for different distributions of effects?  A simple case, when all effects are the same { ($\gamma_i = 1$)}, coincides with Example 2, since the dynamics become the same as for a single locus (Appendix~\ref{apE}). We  consider three distinct distributions of effects to demonstrate both the effect size distributions typically assumed, but also extreme examples. We chose a uniform distribution in $[0,2]$, exponential distribution with mean 1 and a bimodal distribution with many loci of small effect and a few loci of large effect. We chose  all  distributions to have $\langle \gamma \rangle = 1$ for  an easier comparison. 

	\begin{figure}[h]
		\centering
		\includegraphics[width=0.97\textwidth,keepaspectratio]{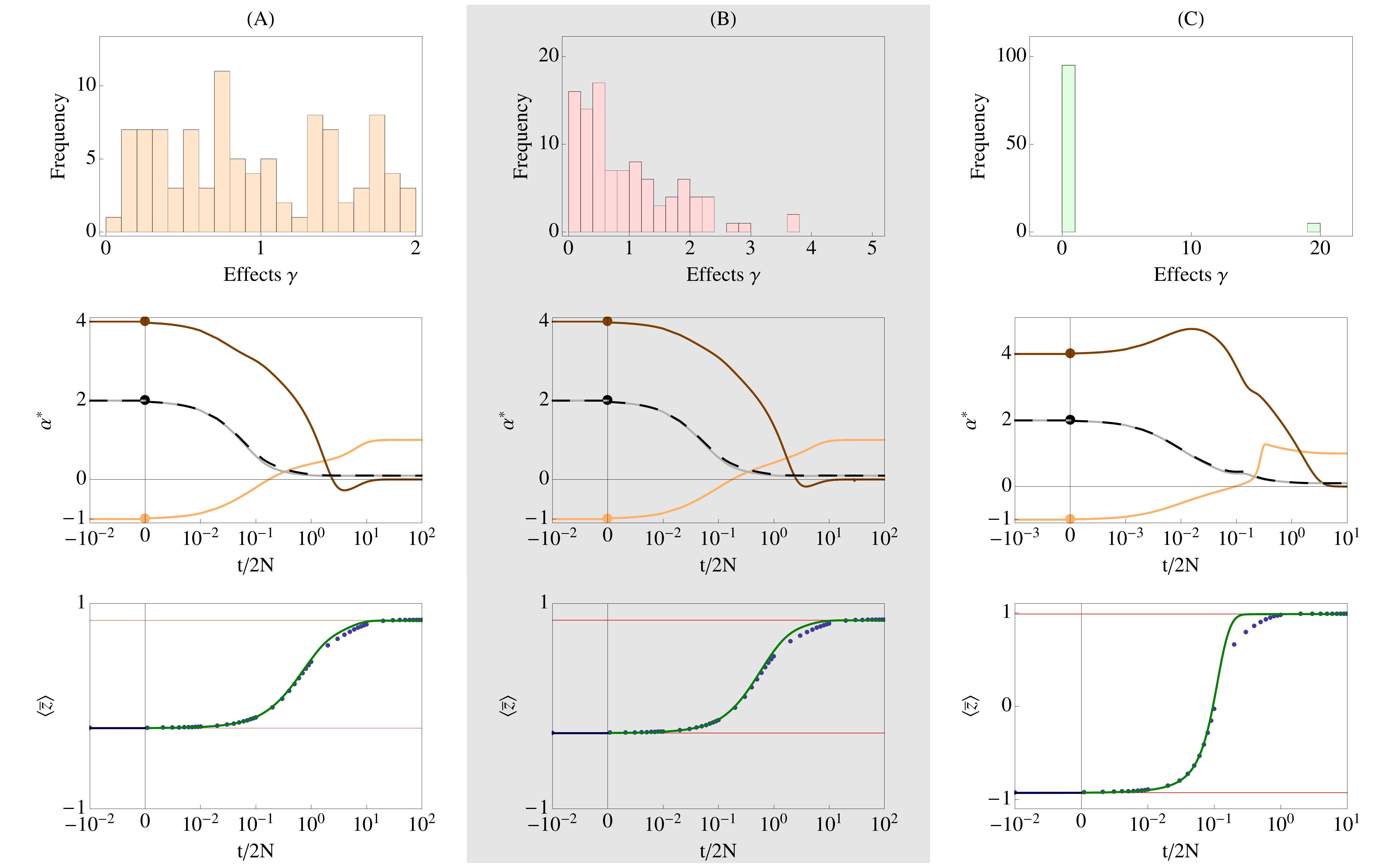}
		\caption{  Example 3, dynamics of  quantitative traits for 100 loci of different effects. 
			The effects are randomly drawn from: (A) uniform distribution in $[0,2]$; (B) exponential distribution with mean 1; and 
			(C) deterministic effects where 95 of the loci have effects 0.01 while the remaining five loci have effects 19.81. 
			The distributions were chosen to have mean equal to one, to be comparable with single-locus simulations. 
			The rapid decrease in mutation is complemented by a change in selection strength and by a change 
			in the selection strength on heterozygosity as in Example 2. 
			{The true forces $\boldsymbol \alpha$ change at time $t=0$ and draw the system out of equilibrium. 
			The response of the trait mean $\langle \bar z \rangle$ is shown in blue dots while 
			the approximation is shown in green. The quality of the approximation for the remaining observables,
			$\langle pq \rangle$, $\langle 2\log p \rangle$, and $\langle 2 \log q \rangle$, can be found in Appendix~\ref{ap_multi}.
			Exact dynamics, obtained by numerically solving the diffusion equation 
			using a discretization of space and time where the explicit transition matrix is known, are
			shown in blue dots. The general DynMaxEnt approximation is shown in green.
			We used $2N=1000$ and $\ep = 1/2N$.
}
		\label{fig_ex3f3} }
	\end{figure}

Figure~\ref{fig_ex3f3} shows a comparison between the exact dynamics and the general DynMaxEnt approximation. All forces, including forward and backward mutation, were initially perturbed to a state where $4N\!\mu = 4N\!\nu<1$, the regime where the continuous approximation does not apply. When the distribution of effects was uniform or exponential, the approximation showed almost a perfect match with the true dynamics in all observed moments. Intuitively, one expects that when the distribution of effects is strongly bimodal, with many small effects and a few large effects, the approximation will perform  badly. This is because the large effect loci dominate the quick response of the system, while the small effect loci form a second wave of adaptation at a later time that is difficult to capture by the simple approximation schemes. 
Figure~\ref{fig_ex3f3} shows these two timescales in the dynamics of Lagrange multipliers. The approximation of the observables still gives a solid match, however, the second wave of adaptation due to small-scale efects is not perfectly accurate anymore.

\section*{Discussion}

A central result in population genetics is that the stationary distribution of allele frequencies is the product of the neutral distribution, and the mean fitness raised to the power of population size: $\psi \sim \phi \bar W^{2N}$  \citep{wright1937,kimura1965,kimura1955stoch}. This result can be interpreted via an optimisation principle: selection constrains the expected values of the traits that determine fitness, but the allele frequency distribution is distorted as little as possible by this constraint -- the distortion being measured by (minus) the relative entropy. 
Our approximation is to assume that this maxent principle holds even away from equilibrium, with the approximate distribution of allele frequencies having the stationary functional \emph{form} at all times. This provides a variational ansatz for the diffusion equation and results in a set of dynamical equations for the parameters of the maxent distribution.

This maximum entropy approximation can be justified in the limit of slowly changing conditions. Yet, provided that mutation rates are above a critical threshold ($4N\mu > 1$), it is remarkably accurate even in the worst case, when parameters change abruptly. Even for a single locus, this approximation gives a substantial reduction in complexity: the whole distribution is described by a small number of dynamical variables that correspond directly to the forces of mutation and selection. The method extends directly to multiple loci, such that the joint distribution of allele frequencies can be approximated by a small number of variables.

This paper extends the approximation to low mutation rates ($4N\mu < 1$), where the original approximation breaks down completely. The problem is that when $4N\mu<1$, populations are likely to be near fixation -- that is, the distribution is concentrated near the borders. When mutation rates or population size change, the distribution near the boundaries changes immediately, whereas the distribution in the interior changes much more slowly. To see why this is so, think of the probability that a population carries one or a few copies of an allele. Because the lifetime of such rare alleles is short, and is determined primarily by mutation and random reproduction, this probability changes rapidly with those processes, and in the short term, is independent of selection or of the total population size. In contrast, the distribution of polymorphic alleles in the interior changes slowly with changes in mutation rate, because it takes a long time for new mutations to reach high frequency, or for polymorphic alleles to be lost by drift.

The maximum entropy approximation for the expectations of a set of traits, $\langle \mathbf A \rangle$, extends the ``breeder's equation'' to include random fluctuations, mutation, and an arbitrary relation between selected traits and the underlying genotype. It takes the form $\partial_t \langle \mathbf A \rangle = { \mathbf B}^\ast (\boldsymbol \alpha-\boldsymbol \alpha^\ast)$ \eqref{BE}, where ${\mathbf B}^\ast$ is a generalisation of the additive genetic covariance, $\boldsymbol \alpha$ are the actual forces of selection and mutation, and $\boldsymbol \alpha^\ast$ are the effective forces that would yield same expectations $\langle \mathbf A \rangle$. Our extension is simply to modify ${ \mathbf B}^\ast$ by truncating the distribution of allele frequencies within a distance $\ep \sim 1/2N$ of the boundaries, thus suppressing its divergence for low mutation rates. The approximation is insensitive to the location of the truncation threshold, $\ep$. We give examples of directional selection and dominance; here, we assume additivity across loci, but \cite{vladar2011stab} show how the method applies to stabilising selection, which induces pairwise epistasis between loci. We emphasise that the approximation can be applied without detailed knowledge of individual loci: only the distribution of allelic effects is needed. The approximation can be also in principle generalized to traits with linkage disequilibrium. The constraints for such a case would also involve pairwise measures, for instance the correlations between loci.

Like the maximum entropy approximation, the infinitesimal model (\cite{fisher1919}, p.403) reduces the dimensionality of the dynamics by following trait values rather than individual alleles. However, these approaches are quite distinct. The infinitesimal model assumes that there are so many genes that the distribution of allele frequencies at each locus is hardly perturbed by selection; thus, the genetic variance that segregates within families remains approximately constant. This is equivalent to assuming that $N\!\beta$ on each allele is small \citep{robertson1960}. In contrast, the maximum entropy approximation can be applied to a single locus, with large $N\!\beta$, and predicts the change in genetic variance due to selection. The infinitesimal model is broader, in that it describes the effects of linkage disequilibrium -- though because it assumes free recombination, these are only significant when selection on traits is strong; see \cite{bulmer1974} for an extension to allow linkage. However, if selection is weak enough that the population is at linkage equilibrium, the maximum entropy approximation will be more accurate than the infinitesimal model, because it accounts for the effects of selection on the genetic variances. Of course, this does require knowledge of the distribution of effects of alleles, and their interactions.

What are the possible applications of our results? 
First, our results allow us to tractably predict the temporal evolution of interesting macroscopic observables, such as the trait mean or heterozygosity, even when these are determined by an arbitrary function of a single locus or multiple loci, under dynamically changing evolutionary forces. This is made possible by the drastic dimensionality reduction of the maximum entropy ansatz. The examples we presented here focused on exploring this ``forward prediction'' scenario in regimes where  previously proposed approximation methods break down. 

The maximum entropy approximation predicts the evolution of the allele frequency distribution - yet almost always, we have only a single realisation of the evolution of any one locus.  However, whole-genome sequencing gives us information about the frequencies of alleles at very large numbers of loci.  If these can be treated as independent realisations of a process common to all loci (or at least, all loci in a functional class), then we can apply our method.  Indeed, our assumptions are the same as those typically made in analysing the distribution of frequencies of synonymous versus non-synonymous variants: each allele is taken to have an additive effect on fitness, drawn from some specific distribution. Usually, the distribution is assumed to be stationary.  However, related species that have different effective population sizes (e.g. \cite{loewe2006}), or newly formed sex chromosomes (e.g. \cite{zhou2013}) require a time-dependent analysis of the kind proposed here.

Second, our results have consequences for inference of evolutionary forces from genomic and phenotypic data. The success of the maximum entropy approximation suggests that the allele frequency distribution remains close to the stationary form, even when selection, mutation and population size are changing rapidly. This in turn suggests that it may be difficult to detect such changes from sequence data taken at a particular time point; note that the moments of the allele frequency distribution correspond directly to the distribution of genealogies. This is consistent with the finding that unless selection is very strong ($N\!\beta \gg1$), it has only weak effects on genealogical structure \citep{williamson2002,barton2004}. 

However, there are several reasons why this conclusion may be too pessimistic. First, even when the maximum entropy approximation accurately predicts the expectations of the observables, the underlying distribution may not necessarily be close to the stationary form (e.g. from \cite{barton2009}, Fig. 10). Second, when $4 N\!\mu < 1$, sudden changes in population size and mutation rate cause immediate changes at the boundaries, so that the distribution deviates from the stationary form -- which is the problem that we address here. Indeed, it is relatively straightforward to detect strong population bottlenecks. Third, specific events, such as the sweep to fixation of a single mutation, can be detected. However, such events occur even when the ensemble is at a stationary state, and so it may still be hard to find evidence for changing conditions. Tests that use an out-group, or examine rates on a phylogeny (e.g. \cite{mcdonald1991,goldman1994,kimura1977}) can detect changes in selection, but require multiple species, and so are not covered by the arguments here. Moreover, our argument from maximum entropy only applies to freely recombining loci: additional information may come from patterns along the genome, which depend on linkage disequilibria and on rates of recombination. Lastly, if instead of data taken at a particular time point, we were provided with the temporal profile of changing observables $\langle\boldsymbol{A}(t)\rangle$, we could use our results to solve an inference problem and learn about  the time courses of evolutionary forces $\boldsymbol {\alpha}$.

Third, our results, together with previous relevant theoretical work, allow us to interpret the evolutionary process in information-theoretic terms.  What is the meaning of entropy, beyond being simply  a tool for approximation? Minus the relative entropy is the Kullback-Leibler distance, a measure of divergence from the neutral distribution. If we include mutation in the base distribution $\phi$ \eqref{neutral}, then minus the relative entropy measures the degree to which selection concentrates populations around states of high fitness. Following \cite{kimura1961} entropy changes can be seen as the information about its selective history that the population can transmit \citep{watkins2008}. Arguably, concentration around fit states is a better measure of adaptation than the increase in mean fitness: though fitness differences determine the rate of adaptation, they do not measure the outcome. Fitnesses may fluctuate, and absolute fitness must stay close to zero on average if populations are to persist.
\cite{mustonen2010} derive an intriguing relation between the gain in information (equal to the reduction in entropy) and the "fitness flux,'' $\Phi$:
	\begin{equation} \label{mustonen}
		2N \mathbb{E} [\Phi ] \ge - \Delta S \quad \text{where} \quad \Phi = \int_0^T \sum_i \beta_i \Delta p_i dt
	\end{equation}
(\eqref{mustonen} corrects a factor 2 error in \cite{mustonen2010}). This applies to a haploid population at linkage equilibrium, as assumed throughout this paper. Selection may change arbitrarily over time, so that this relation gives a lower bound on the fitness flux, $\Phi$, that is required to achieve a given gain in information (i.e., reduction in entropy, $\Delta S$). If selection changes slowly -- as required for our MaxEnt approximation to be accurate -- then the inequality approaches an equality. The fitness flux can be separated into a component due to selection (which must be positive and equal to the additive variance in fitness, $V_W$), and the remaining components, due to mutation and drift.
Because forces other than selection are expected on average to act against adaptation, the latter component is negative, so that the additive variance in fitness should set a bound on the rate of information gain (i.e. $2NV_W \ge -\Delta S$).

Our approximation states that even out of equilibrium, the distribution of allele frequencies minimises the information gain, subject to constraints on selected traits. By drastically reducing the dimensionality of the system, to cover only the expectations of selected quantities, we can simplify expressions for the total fitness flux and variance in fitness over the evolutionary trajectory, and may therefore be able to understand how these quantities limit the amount of information that can be accumulated by selection.

\section*{Acknowledgements}
The authors thank Harold de Vladar and Richard Koll\'ar for helpful discussions. 
The research leading to these results has received funding from the European Research Council under the European Union's Seventh Framework Programme (FP7/2007-2013) / ERC grant agreement Nr. 250152 (NB).
This work was supported in part by Human Frontiers Science Program Nr. RGP0065/2012 (GT).


\appendix
\setcounter{table}{0}
\setcounter{figure}{0}
\numberwithin{equation}{section}
\numberwithin{figure}{section}

\section{Solution of the MaxEnt by the method of Lagrange multipliers} \label{apLM}
The variational problem
	\begin{align}
		\max_{\psi} S_H[\psi] &= -\int_{[0,1]^{ L}} \psi(\mathbf{p},t)\log\left[ \frac{\psi(\mathbf{p},t)}{\phi(\mathbf{p})}\right] d\mathbf{p}  
	\end{align} 
with $\phi(\mathbf p) = \prod_{i=1}^L (p_i q_i)^{-1}$ subject to constraints 
	\begin{align}
		\int_{[0,1]^L}  \psi(\mathbf p) d\mathbf p  = 1\,, \quad \int_{[0,1]^L}  \mathbf A(p) \psi(\mathbf p) d\mathbf p  = \langle \mathbf A \rangle_\text{obs}\,,
	\end{align} 
(for example for directional selection and asymmetric mutation $\mathbf A = (\bar z, U,V)$ and $\boldsymbol \alpha = (\beta, \mu, \nu)$) can be solved by the method of Lagrange multipliers, yielding an unconstrained maximization of a Lagrangian 
	\begin{align}
		\mathcal{L}[\psi,\lambda,\alpha] &= S_H[\psi] - \lambda\left[ \int_{[0,1]^L}  \psi(\mathbf p) d\mathbf p  -1\right]
			- 2N \sum_k  \alpha_k \left[ \int_{[0,1]^L}  A _k\psi(\mathbf p) d\mathbf p 
			 - \langle \mathbf A_k \rangle_\text{obs} \right]
	\end{align}
with multipliers $\lambda$ and $\boldsymbol\alpha$.
The variational derivative of this function with respect to its argument $\psi$ is
	\begin{align}
		\delta \mathcal{L}[\psi,\lambda,\alpha] &=   \int_{[0,1]^L}  \delta \psi \left[ \log \frac{\psi}{\phi} 
		-1-\lambda - 2N\boldsymbol \alpha \cdot \mathbf A \right] d\mathbf p 	 
	\end{align}
and leads to a solution
	\begin{align}
		\bar \psi(\mathbf p)&=  \frac{\phi(p)}{\mathbb{Z}} e^{2N\boldsymbol \alpha \cdot \mathbf A}
	\end{align}
where the normalization $\mathbb{Z} = \exp(-1-\lambda)$ and the Lagrange multipliers are such that observables are correctly matched. The distribution $\bar \psi(\mathbf p)$ coincides with the stationary solution of the Fokker-Planck equation \eqref{FPE} provided $\boldsymbol \alpha$ are the evolutionary forces (selection, mutation) and $\mathbf A$ are the traits associated with the underlying processes. The evolutionary forces appear in the constrained optimization as Lagrange multipliers $2N\boldsymbol \alpha$.

\section{Matrices $\mathbf B$ and $\mathbf C$ in the continuous DynMaxEnt approximation} \label{apBC}
When $\boldsymbol \alpha = (\beta, h, \mu, \nu)$ and $\mathbf A = (\bar z, H, U, V)$, the explicit form of matrix $\mathbf B$ is
	 \begin{align} \label{Bexplicit}
		\mathbf B^\ast &=\left  \langle \sum_{k=1}^L \left(
		\begin{array}{ccccc}
			\gamma_k^2 p_k q_k&2\gamma_k p^2_k q_k (q_k-p_k)&\gamma_k q_k & -\gamma_k p_k\\ 
			2\gamma_k p^2_k q_k (q_k-p_k)& 2p_k^3 q_k (q_k-p_k)^2&2p_k q_k (q_k-p_k) &-2p_k^2 (q_k-p_k)\\
			\gamma_k q_k &2p_k q_k (q_k-p_k) &2\frac{q_k}{p_k}&-2\\
			-\gamma_k p_k & -2p_k^2 (q_k-p_k) &-2&2\frac{p_k}{q_k}
		\end{array}
		\right) \right \rangle \,.
	\end{align}
Since the expectation 
$$\langle f(\mathbf p) \rangle = \int_{[0,1]^L} f(\mathbf p) \bar \psi (\mathbf p) d\mathbf p$$ 
is taken over the stationary distribution $\bar \psi(\mathbf p)$ that depends on the effective parameters $\boldsymbol \alpha^\ast$, the matrix $\mathbf B^\ast$ is also a function of the effective evolutionary forces. The components of matrix $\mathbf B^\ast$ may be calculated using special functions as in \cite{barton2009}.

Similarly, the components of the matrix $\mathbf C^\ast$ can be written as
	\begin{align*}
		\boldsymbol C_{i,j}^\ast &=\text{Cov}(A_i,A_j) = \langle  A_iA_j \rangle - \langle  A_i \rangle
		\langle  A_j \rangle \,, 
	\end{align*}
$A_i(\mathbf p)$ are functions of the microscopic allele frequencies and after averaging over the stationary distribution $\bar \psi(\mathbf p)$
only the dependence on the effective forces $\boldsymbol \alpha^\ast$ remains. Thus the right hand side of the vector equation \eqref{QSDA} is solely a function of effective evolutionary forces $\boldsymbol \alpha^\ast$, forming a system of ordinary differential equations of dimension equal to the number of observables.

\section{Discrete dynamics in the limit of small $N\!\mu$} \label{apD}

As $4N\!\mu$, $4N\!\nu$ become very small, the probability distribution \eqref{stationary} becomes concentrated at the boundaries. The populations then switch between { fixation for the favourable and deleterious  alleles at a given locus and can be described by the fraction $P$, { $Q=1-P$}, of populations fixed (or nearly fixed) for each allele}
	\begin{align}
		\frac{dP}{dt}&= \lambda_+ Q - \lambda_- P\,.
	\end{align}
The probability of fixation for the favourable allele is $\hat P  = (\nu/\mu + \exp(-4N\!\beta))^{-1}$ \citep{kimura1962} and the rates of substitutions of alleles by their counterparts, $\lambda_+$ and $\lambda_-$, are 
	\begin{align}
		\lambda_- = 4N\!\nu \beta \frac{e^{-4N\!\beta}}{1-e^{-4N\!\beta}}\,, \qquad \lambda_+ 
		= 4N\!\mu \beta \frac{1}{1-e^{-4N\!\beta}}\,.
	\end{align}
Hence, the exact dynamics in the regime of small mutation have a form:
	\begin{align} \label{discrete}
		4N\frac{dP}{dt}&=  4N\! \beta \frac{4N\!\mu Q-4N\!\nu e^{-4N\!\beta}P}{1-e^{-4N\!\beta}}\,.
	\end{align}

How  does the standard MaxEnt approximation compare, in the limit of low mutation rates? We keep mutation rates fixed, and follow a single variable, $P=\langle p \rangle$; we define the complementary variable $\beta^\ast$ as the selection that gives $P$ at stationarity.  In the limit of low mutation rates:
	\begin{align}
		P&=  \frac{\int_0^1 p\cdot p^{4N\!\mu-1}q^{4N\!\nu-1}e^{4N\!\beta^\ast { (2p-1)}}dp}{\int_0^1  
		p^{4N\!\mu-1}q^{4N\!\nu-1}e^{4N\!\beta^\ast { (2p-1)}}dp}
		\approx \frac{\frac{1}{4N\!\nu}}{e^{-4N\!\beta^\ast} \frac{1}{4N\!\mu}+\frac{1}{4N\!\nu}}\,,
	\end{align}
implying
	\begin{align}
		e^{4N\!\beta^\ast} &= \frac{4N\!\nu}{4N\!\mu} \frac{P}{Q}\,.
	\end{align}
The same equilibrium formula is given by the ratio of substitution rates, $\frac{P}{Q}=\frac{\lambda_+}{\lambda_-}$.  Under the MaxEnt approximation, the rate of change is:
	\begin{align} \label{dynP1}
		4N\frac{dP}{dt}&= \langle pq \rangle (4N\!\beta-4N\!\beta^\ast) \,.
	\end{align}
Moreover, in the limit of small mutation
	\begin{align}
		\langle pq \rangle &=  \frac{\int_0^1  p^{4N\!\mu}q^{4N\!\nu}e^{4N\!\beta^\ast { (2p-1)}}dp}{\int_0^1  
		p^{4N\!\mu-1}q^{4N\!\nu-1}e^{4N\!\beta^\ast { (2p-1)}}dp}  
		\approx \frac{e^{4N\!\beta^\ast}-1}{4N\!\beta^\ast \left( \frac{1}{4N\!\mu} 
		+ \frac{e^{4N\! \beta^\ast }}{4N\!\nu} \right)}\,, \label{pq}
	\end{align}
{where the above expression has been obtained by computing contributions to the integrals separately for each of the boundaries $[0,\delta]$ and $[1-\delta,1]$ for $\delta \ll1$ and for $4N\!\mu, 4N\!\nu \ll \delta$. Note that integration by parts has been used, resulting in the presence of the term $4N\!\beta^\ast$ in the denominator of \eqref{pq}.}
Equation \eqref{dynP1} then becomes
	\begin{align} \label{maxent_smallmut}
		4N\frac{dP}{dt}&=  \frac{4N\!\nu P-4N\!\mu Q}{4N\!\beta^\ast} (4N\!\beta-4N\!\beta^\ast) 
		=\left[ \frac{4N\!\beta}{\log\left( \frac{4N\!\nu}{4N\!\mu} \frac{P}{Q}\right)}-1 \right] ( 4N\!\nu P-4N\!\mu Q )\,.
	\end{align}
Thus, the MaxEnt approximation to the full distribution does not converge to the exact dynamics \eqref{discrete} as mutation rates become small.  Nevertheless, the dynamics are approximated quite well, provided that selection is not too strong (e.g., $|4N\!\beta|<2$).  
The MaxEnt approximation greatly underestimates the rate of change at the margins, and gives no effect of selection at the exteme allele frequencies, see Figure~\ref{fig_discrete}, left (all alphabetical figure citations refer to figures in the appendices). 

	\begin{figure}[h]
		\centering
		\includegraphics[width=0.77\textwidth,keepaspectratio]{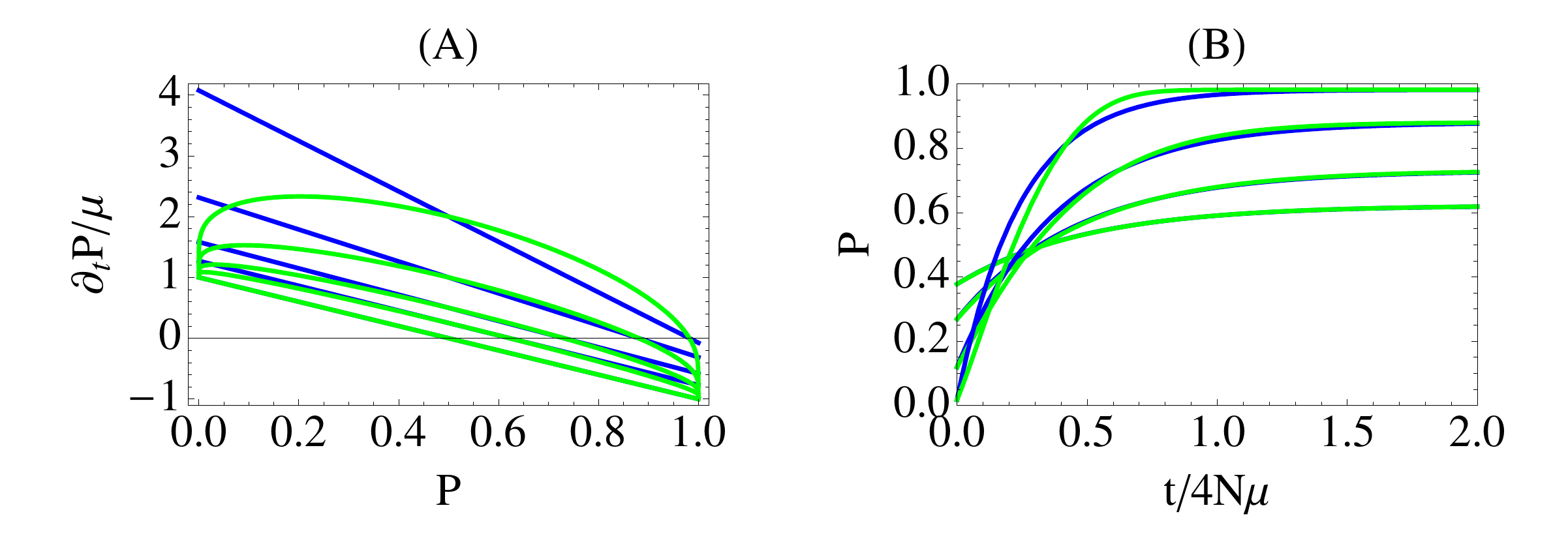}
		\caption{\label{fig_discrete}
			Discrete MaxEnt approximation.
			(A) The exact rate of change, obtained from the discrete model \eqref{discrete} (blue), compared with the MaxEnt approximation \eqref{maxent_smallmut} (green), 
			in the limit of small $4N\!\mu=4N\!\nu$; $N\!\beta=0,0.5,1,2, 4$ (bottom to top) and $2N=1000$.  
			Rates are scaled relative to the neutral mutation rate, assumed symmetric ($\mu=\nu$).  
			(B) Dynamics following an abrupt switch from $4N\! \beta = -0.5, -1, -2, -4$ to the reverse.  }
		\label{fig1}
	\end{figure}

However, the equilibria necessarily agree: the exact and the approximate rates of change are zero at the same point.  The right plot in Figure~\ref{fig_discrete} shows the exact $P(t)$, and the MaxEnt approximation; these are close up to $4N\! \beta=\pm 2$, but the solution becomes poor for $4N\! \beta=4$ (upper pair of curves).  The MaxEnt solution is accurate for $4N\! \beta\le 2$ because $P$ remains within the interior ($\frac{1}{1+e^2}=0.112<P<0.888$), where the rate of change is approximated well.

\section{Failure of the continuous DynMaxEnt for $4N\!\mu, 4N\!\nu < 1$} \label{apC}

The log mean fitness will typically be a sum over moments of allele frequencies.  For example, a selection gradient $\beta$ on a trait with mean { $\bar z = 2\sum_i \gamma_i (p_i-q_i)$} will introduce a component where  $\boldsymbol \alpha_W = \beta$ and $\mathbf A_W = \bar z$; model with dominance { requires $\boldsymbol \alpha_W = (\beta,h)$ and $\mathbf A_W = (\bar z,H)$} and epistasis introduces mixed second order moments of the allele frequencies.  Thus, the matrix ${\mathbf B}$ is an expectation over polynomial functions of allele frequencies, and is well-behaved.

In contrast, the elements of ${\mathbf B}$ that describe mutation diverge when $4N\!\mu<1$ or $4N\! \nu<1$.  To see this, consider a single locus for which  $\mathbf A = (\bar z, { H}, U, V) = (2p-1, { 2pq,} 2\log p, 2\log q)$, and the elements { $B_{3,3}$, $B_{3,4}$, $B_{4,4}$} are $\langle 2q/p\rangle$, $\langle -2 \rangle$, $\langle 2p/q\rangle$ \eqref{Bexplicit}.  Thus, { $B_{3,3}$} diverges when $4N \mu<1$, and {$B_{4,4}$} diverges when $4N\nu<1$.  If we can assume that mutation rates are fixed, then we can avoid the difficulty either by fixing the mutation rates always at their actual values (i.e., $\mu=\mu^\ast$, $\nu=\nu^\ast$), or by choosing a reference distribution that includes mutation: $\phi=\prod_k p_k^{4N\!\mu-1} q_k^{4N\!\nu-1}$, and dropping the observables $U$, $V$.  These two approaches are equivalent since fixing the mutation rate leads to
{ $\partial_t \langle \bar z \rangle= B_{1,1}(\beta-\beta^\ast) + B_{1,2}(h-h^\ast) $} leading to the same dynamics for { $\beta^\ast$ and $h^\ast$} as if the reference distribution included mutation.

\begin{figure}[h]
		\centering
		\includegraphics[width=0.98\textwidth,keepaspectratio]{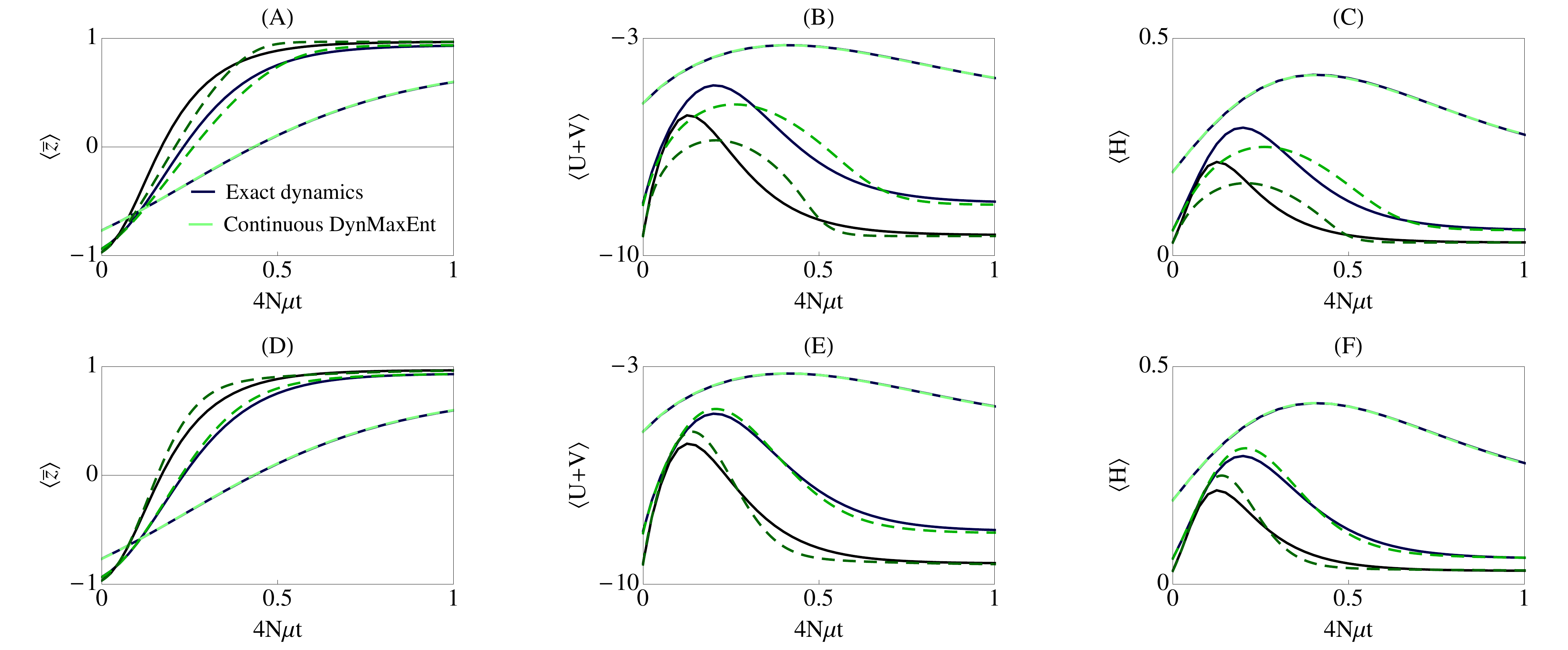}
		\caption{\label{figC1}
			Failure of the continuous DynMaxEnt approximation for strong selection.
			Dynamical response to a fast change in selection $N\!\beta =-4$ to $N\!\beta =4$ for symmetric mutation rates that
			are unperturbed and $2N=1000$. 
			Each panel shows three simulations with mutation rates $4N\!\mu=4N\!\nu \in \{2,1/2,1/4\}$ (light-green to dark-green)
			where the exact observables (black) are compared with the  continuous DynMaxEnt
			approximation (green, dashed).
			(A-C) Simulation with constraints $\langle\bar z\rangle$, $\langle U \rangle$, $\langle V \rangle$ 
			when $4N\!\mu,4N\!\nu>1$ 
			and a single constraint $\langle \bar z \rangle$ when $4N\!\mu,4N\!\nu<1$.  
			(D-F) Simulation with constraints $\bar z$, $\langle H \rangle$, $\langle U \rangle$, 
			$\langle V \rangle$  when $4N\!\mu,4N\!\nu>1$ 
			and two constraints $\langle \bar z \rangle$, $\langle H \rangle$ when $4N\!\mu,4N\!\nu<1$.} 
	\end{figure}

We first explore the  continuous DynMaxEnt approximation for $4N\!\mu>1$ and compare its accuracy with $4N\!\mu<1$. We study the worst-case scenario when a selection  suddenly changes sharply from $N\!\beta = -4$ to $N\!\beta = 4$; this is by no means an adiabatic change. Mutation equals $4N\!\mu = 4N\!\nu = 2,1/2,1/4$ where the first choice ($4N\!\mu>1$) allows a full  continuous DynMaxEnt approximation including constraints on $\langle U\rangle$ and $\langle V\rangle$ while the second and third choices require a fixed instantanous mutation rate and constraints on $\langle U \rangle $ and $\langle V \rangle$ dropped to keep the entries of the additive covariance matrix $B$ finite.  Figure~\ref{figC1} (top row) shows the predicted observables $\langle \bar z\rangle$, $\langle U+V\rangle$ and $\langle H\rangle$ estimated by the continuous DynMaxEnt for each of the mutation rate alternatives compared with the exact solution, { while keeping the heterozygosity fixed throughout the simulation (not employing the constraint on $\langle 2pq \rangle$)}. The  method is accurate for $4N\!\mu=2$ (where also the mutation changes dynamically) but shows significant deviations from the true dynamics for $4N\!\mu = 4N\!\nu = 1/2, 1/4$. 

{ We also show simulations (D-F) where the selection on heterozygosity $h$ is treated as a dynamical force, employing an additional constraint on $\langle H \rangle$. This increases the number of degrees of freedom in the approximate dynamics and affects the accuracy of the approximation. For instance, the decrease in performance for $4N\!\mu,\, 4N\!\nu <1$ is partially caused by losing two degrees of freedom by fixing $\mu$ and $\nu$. On the other hand, increase in accuracy in Figure~\ref{figC1} (bottom row) is caused by introducing additional constraints on $\langle 2pq\rangle$. }  This is visible both for $4N\!\mu  >1$ where the number of Lagrange multipliers was increased from three to four and for $4N\!\mu<1$ where the constraint on $\langle \bar z \rangle$ was complemented by the second constraint on a mean heterozygosity $\langle H \rangle = \langle 2pq \rangle$.

\begin{figure}[h]
		\centering
		\includegraphics[width=0.98\textwidth,keepaspectratio]{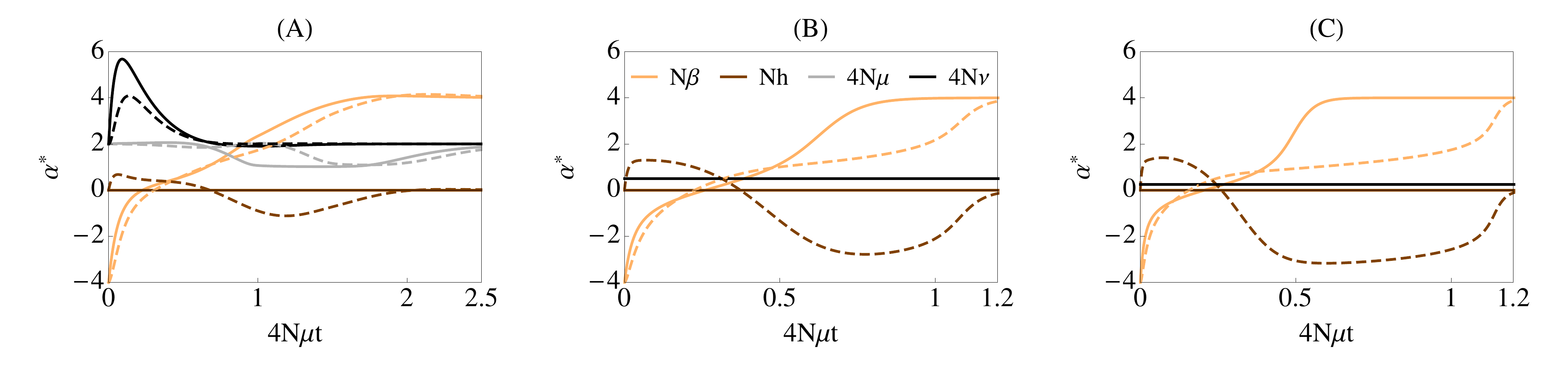}
		\caption{\label{figC2}
			Effective coefficients $\boldsymbol \alpha^\ast$ of the continuous DynMaxEnt approximation for scenario of 
			Figure~\ref{figC1}. 
			(A) $4N\!\mu = 4N\!\nu = 2$,
			(B) $4N\!\mu = 4N\!\nu = 1/2$, and
			(C) $4N\!\mu = 4N\!\nu = 1/4$ are compared between the simulation  without (solid) versus with (dashed) the
			heterozygosity as a degree of freedom. The population size is $2N=1000$. }
	\end{figure}

Figure~\ref{figC2} presents the dynamics of Lagrange multipliers in the example in Figure~\ref{figC1} where a sudden change in selection  from $N\!\beta = -4$ to $N\!\beta = 4$ was applied to  three systems, with different mutation rate $N\!\mu,\, N\!\nu \in \{ 2,1/2,1/4\}$. While in the superthreshold regime $4N\!\mu,\, 4N\!\nu >1$, mutation is allowed to change in the continuous DynMaxEnt approximation, in the subthreshold regime $4N\!\mu,\,4N\!\nu <1$ it is fixed. 

Figure~\ref{figC2} suggests that 
increasing the number of constraints; which increases the dimensionality of the problem, is associated with a slower convergence to a steady-state and a separation of timescales. This is visible both for $4N\!\mu  >1$ (left) where the number of Lagrange multipliers was increased from three to four and for $4N\!\mu<1$ where the constraint on $\langle \bar z \rangle$ was complemented by the second constraint on a mean heterozygosity $\langle H \rangle = \langle 2pq \rangle$.

Next, suppose that the scaled mutation rate is initially high enough that the stationary distribution is concentrated in the interior, but then abruptly falls below the threshold at which populations are typically near fixation: $4N\!\mu=4N \!\nu$ falls from 2 to 0.1.  (Note that a fall in $N\!\mu$ could also be due to a fall in population size rather than in mutation rate). We also assume that selection changes abruptly at the same time, i.e., $N \!\beta$ changes from $-1$ to $1$, in order to give the most challenging example: errors in estimating $\langle pq \rangle$ will be reflected in errors in the rate of change of the mean.  Immediately after the fall in scaled mutation rate, probability accumulates at the boundary and develops a boundary layer of a form $\sim p^{4N\!\mu-1}q^{4N\!\nu-1}$ that agrees with the stationary form, but polymorphism decays more slowly so the full distribution is not from the stationary class.  The continuous DynMaxEnt  fails to capture the true dynamics since the mutation rates have to be instantly adjusted to the terminal values.

	\begin{figure}[h]
		\centering
		\includegraphics[width=0.98\textwidth,keepaspectratio]{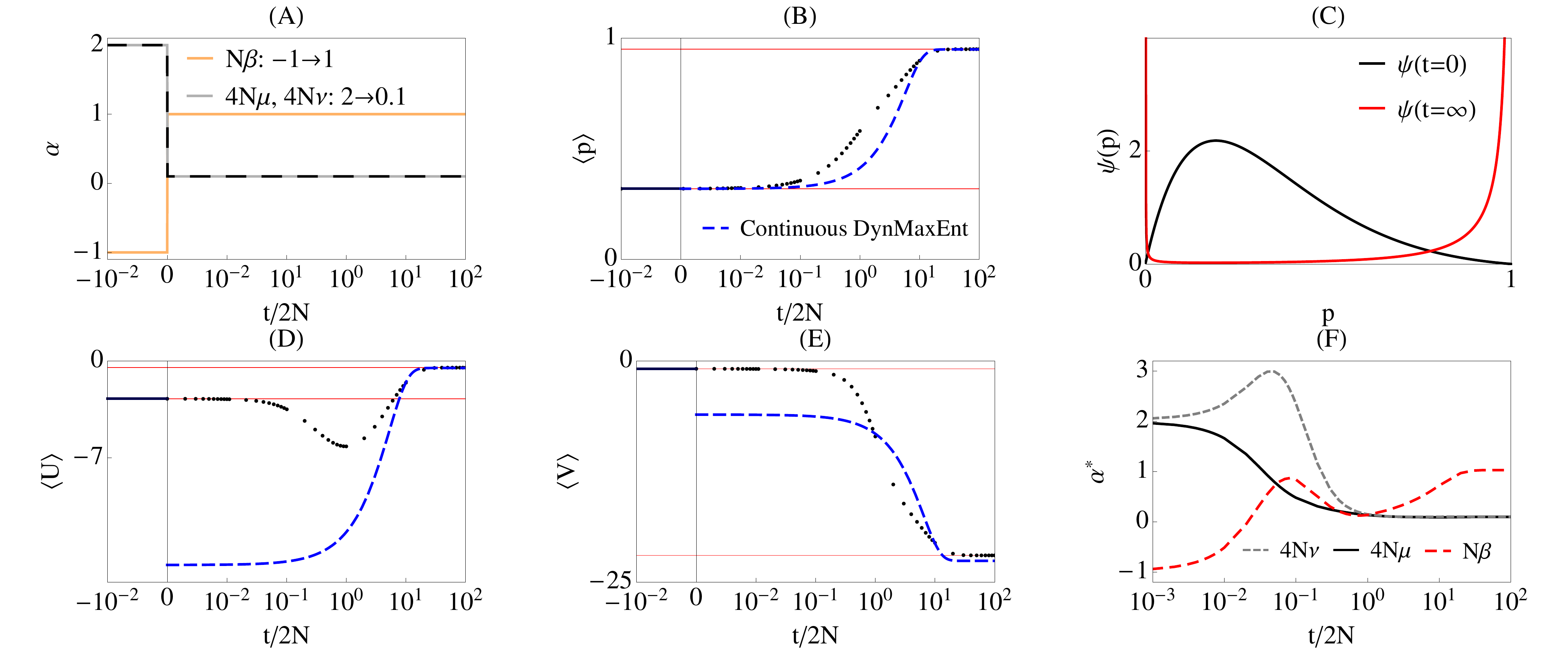}
		\caption{\label{figC3} 
			Failure of continuous DynMaxEnt approximation for changing mutation $4N\!\mu<1$  and $2N=1000$.
			(A) Changes in evolutionary forces  that draw the system out of equilibrium
			include a rapid decrease in mutation, complemented 
			by a change in selection strength.
			(B,D,E) The response of the observed quantities $\langle p \rangle$,  { $\langle U \rangle$ and 
			$\langle V \rangle$.}
			(C) The initial and final stationary allele frequency distributions.
			(F) The effective forces  $\boldsymbol \alpha^\ast$.}
	\end{figure}

Figure~\ref{figC3} shows that three measures of diversity ($\langle U \rangle, \langle V\rangle, \langle pq \rangle$) change rapidly, with $\langle U \rangle, \langle V\rangle$ falling most rapidly because they are more sensitive to the rapid changes near the boundary than $\langle pq \rangle$.  Note that $\langle U \rangle = \langle 2\log p \rangle$ falls until $t\sim {1}/{2N}$, as probability accumulates close to $p=0$ (note the log scale on Figure~\ref{figC3}), but then increases again more slowly, as favourable mutations substitute, transferring probability away from $p\sim 0$.  The mean changes rapidly and substantially while heterozygosity is still high, but then changes more slowly after $t\sim 1/2N$, when the genetic variance is low, and selection is limited by the influx of new mutations. 

The changes in effective parameters $\boldsymbol \alpha^\ast$ are shown in Figure~\ref{figC3} (F).  Because we assume that the population is initially in its stationary state, these necessarily begin at their actual values ($4N \!\mu^\ast$, $4N\! \nu^\ast = 2$, $N\! \beta^\ast=-1/2$).  After the mutation rate decreases, the probability of being fixed increases rapidly, and therefore $4N \!\mu^\ast$, $4N \!\nu^\ast$ fall quickly, approaching their new values by $t/2N\sim 0.1$; $\mu^\ast$ falls first, because the probability of being near $p=0$ increases faster than the probability of being near $p=1$.  Over this time, $N\! \beta^\ast$ changes even while the mean $\langle p \rangle$ hardly changes, in order to compensate for changing $N\!\mu$, $N\!\nu$.  At later times, when the effective mutation rates are constant and close to their actual (low) rates, $N \!\beta^\ast$ increases as the mean changes.

Fixing the effective mutation rates { makes DynMaxEnt unable} to capture transient properties of the adaptation process. The dynamics of the mean allele frequency follows $p' = 2\langle pq \rangle (\beta-\beta^\ast)$ where $\beta^\ast$ can be obtained at each time as a value that gives the current mean allele frequency given $\mu^\ast = \mu$ and $\nu^\ast = \nu$. { The effective} directional selection can be further used to compute the change of observables, as displayed in Figure~\ref{figC3}  (dashed blue). This continuous DynMaxEnt approximation fails to capture transient dynamics of the observed quantities but still converges to the correct state. If the exact dynamics are followed for a short time from the switch of the evolutionary forces and only then is the continuous DynMaxEnt approximation initialized, the approximation would more closely agree with the actual observables. This is because the real dynamics of the trait mean slows down in time as the system gets close to an equilibrium and therefore is better captured by an approximate process, whose speed of adaptation is limited by the small fixed mutation rate.

\section{Derivation of the general DynMaxEnt approximation for low mutation { from the diffusion equation \eqref{FPE}}} \label{apDerivation}

We consider a directional selection acting on a single locus. The extension to multiple loci with different effects and overdominance is straightforward and discussed later. 

The failure of the continuous DynMaxEnt method, described in Appendix~\ref{apD} and \ref{apC} requires a special treatment of the boundaries of the allele frequancy domain to avoid divergence of the method due to singularities in matrix $B$. A naive resolution is  to completely disregard the boundaries, and to simply truncate the allele frequency distribution. This ad hoc method, despite resolving problems with divergence of matrix $B$, would not capture the dynamics at the boundaries very accurately, particularly in the small mutation regime. Thus we approach the problem in a more pedantic way. 

First, we split the domain to the interior and the boundaries and propose a dynamical model that captures all essential processes. The new system depends on a truncation parameter $\ep$, representing the width of the boundary layer. The stationary distribution of the system can still be represented as a solution of the variational problem with two additional constraints and Lagrange multipliers, related to the boundary masses. The DynMaxEnt method can be derived for this expanded system. However, the resulting macroscopic dynamics of the effective forces involves inclusion of the coupling terms at $p=\ep, 1-\ep$ which are microscopic quantities; these are not in general accessible. Therefore, we introduce an approximation that does not directly prescribe the probability fluxes but circumvents this by setting the total mass at the boundaries consistently with the continuous stationary distribution. This step is crucial for our approximation. The effect is that the expanded dynamics is brought back to a dimension where only the mutation and selection forces are followed.

The derivation of the general DynMaxEnt approximation for a system where the stationary solution is a discontinuous function 
	\begin{align}\label{PDE2sol}
		\psi_\ep (p,t) = 
		\begin{cases}
			\psi^1_\ep(p,t) & p<\ep\\
			\psi^2_\ep(p,t) & \text{else}\\
			\psi^3_\ep(p,t) & p>1-\ep
		\end{cases}
	\end{align}
consists of the following steps:
	\begin{itemize}
		\item[(a)]
		Set up a piecewise-defined diffusion dynamics with the appropriate stationary solution.
		\item[(b)]
		Use the stationary solution as an ansatz for a solution of the diffusion problem allowing only the 
		Lagrange multipliers to change in time.
		\item[(c)]
		Derive dynamics of the means of observables from the above system and reduce it to a closed 
		dynamical system for Lagrange multipliers.
	\end{itemize}

\noindent(a) 
{
Diffusion equation \eqref{FPE} can be written in the flux form
	\begin{align} 
		\frac{\partial \psi}{\partial t} & =  -\frac{\partial}{\partial p} J[\psi,p ]  
	\end{align}
where the flux $J$ is defined as 
	\begin{equation} \label{J}
		J[\psi,p ] = 
			 \left[ \mathcal{M}_i(p_i) \psi \right]
			- \frac{1}{2}  \frac{\partial}{\partial {p}} \left[ \mathcal{V}_i(p_i) \psi \right]\,.
	\end{equation}
We set up the dynamical system for $\psi_\ep (p,t)$ separately in the three regions of the domain:
	\begin{align} \label{PDE2}
		\frac{\partial \psi_\ep}{\partial t} & =  -\frac{\partial}{\partial p} J_\ep[\psi_\ep,p ]   = 
		\begin{cases}
			-\frac{\partial}{\partial p} J_\ep^1[\psi_\ep^1,p ]  & p\in [0,\ep)\\
			-\frac{\partial}{\partial p} J_\ep^2[\psi_\ep^2,p ] & p\in (\ep,1-\ep)\\
			-\frac{\partial}{\partial p} J_\ep^3[\psi_\ep^3,p ]  & p\in (1-\ep,1]
		\end{cases}
	\end{align}
where $J_\ep$ are the fluxes in the system with split domain \eqref{fluxes2}
	\begin{align} \label{fluxes2}
		\begin{cases}
			J_\ep^1[\psi^1,p ] = 
			 \left[ (\beta \ep q -\nu \ep) \psi^1_\ep \right]
			- \frac{1}{4N}  \frac{\partial}{\partial {p}} \left[ \ep q \psi^1_\ep \right] \\
			J_\ep^2[\psi^2,p ] = 
			\left[ (\beta pq -\nu p + \mu q) \psi^2_\ep \right]
			- \frac{1}{4N}  \frac{\partial}{\partial {p}} \left[ pq \psi^2_\ep \right] \\
			J_\ep^3[\psi^3,p ] = 
			\left[ (\beta p\ep + \mu \ep) \psi^3_\ep \right]
			- \frac{1}{4N}  \frac{\partial}{\partial {p}} \left[ p \ep \psi^3_\ep \right] 
		\end{cases}
	\end{align}
and where the coupling between regions, which influences the probability of extreme allele frequencies, is set as
	\begin{eqnarray}
		J^1_\ep[\psi^1_\ep,0] &=&    0 \label{J1}\\ 
		J^3_\ep[\psi^3_\ep,1] &=&  0 \label{J2}\\
		J^1_\ep[\psi^1_\ep,\ep] &=& J^2_\ep[\psi^2_\ep,\ep] =  J[\psi,\ep] \label{J3} \\
		J^2_\ep[\psi^2_\ep,1-\ep] &=& J^3_\ep[\psi^3_\ep,1-\ep] = J[\psi,1-\ep] \label{J4}
	\end{eqnarray}
}
While the conditions \eqref{J1}-\eqref{J2} at $p=0,1$ ensure that the probability mass does not dissipate through the boundaries, the conditions at $p=\ep,1-\ep$ set the fluxes in accordance with the fluxes of the original continuous dynamics \eqref{FPE} which are not explicitly known. This results in the same stationary solution in the bulk as in the original diffusion process. Moreover, the probability masses at the boundaries will also agree  with the standard problem since the fluxes are identical.
The stationary solution, { computed for $p\in [0,\ep)$ as a solution of the problem $J^1_\ep[\psi^1,p] = 0$, and analogously for  subdomains $(\ep,1-\ep)$ and $(1-\ep,1]$, has a form \eqref{SCdist2}. Moreover, setting the total mass at each subdomain equal to the corresponding mass of the stationary solution \eqref{stationary} to leading order implies the choice of constant prefactors in \eqref{SCdist4}.}

\vskip 0.3cm
\noindent(b) The stationary distribution $\bar \psi_\ep(p|\boldsymbol \alpha)$ in \eqref{SCdist2} depends on five parameters (for directional selection), including $\kappa$ and $\rho$ related to boundary masses. These parameters relate to a set of observables $\mathbf A^\text{full} = ( \bar z, U_\ep , V_{\ep} ,\chi_{[0,\ep]} , \chi_{[1-\ep,1]} )$ as the corresponding Lagrange multipliers $2N\boldsymbol \alpha^\text{full} = 2N\!\left(\!\beta ,\mu,\nu,\kappa,\rho \right) $. We derive the full DynMaxEnt method for this extended problem. Only after the derivation of the full dynamics will we incorporate the relationship between $\kappa,\mu$ and $\rho,\nu$. We use an ansatz 
	\begin{align}
		\psi_\ep(p,t) = \bar \psi_\ep(p|\boldsymbol \alpha^\ast) + \delta(p,t)
	\end{align}
for the the solution of \eqref{PDE2} that expresses the non-stationary solution of the equation as a sum of a stationary distribution, with effective Lagrange multipliers $\boldsymbol \alpha^\ast$ changing in time, and $\delta$, representing the deviation from the stationary form. 
First we express the left-hand side of the \eqref{PDE2} using a chain rule:
	\begin{eqnarray*}
		\frac{\partial  \bar\psi_\ep}{\partial t} &=& \frac{\partial  \bar\psi_\ep}{\partial \beta^\ast} \frac{\partial \beta^\ast}{\partial t}
			 +\dots + 
			\frac{\partial  \bar\psi_\ep}{d\rho^\ast} \frac{\partial \rho^\ast}{\partial t} 
			+ \frac{\partial \delta}{\partial t}\,.
	\end{eqnarray*}
Next we calculate each summand and for clarity drop the notation in the exponent of $\boldsymbol \alpha^\text{full}$
	\begin{eqnarray*}
		 \frac{\partial  \bar\psi_\ep}{\partial \alpha_k^{\ast }} &=&
		-\frac{1}{\mathbb Z} \frac{\partial \mathbb Z}{\partial \alpha_k^\ast} \bar \psi_\ep + 2N A_k \bar \psi_\ep 
		= 2N [ A_k - \langle A_k \rangle  ] \bar \psi_\ep\,,
	\end{eqnarray*}
where $\langle A_k \rangle$ arises from differentiating $\mathbb Z$, and $A_k$ from differentiating $e^{2N \boldsymbol \alpha \cdot \mathbf A}$ with respect to $\alpha_k^\ast$. This leads to
	\begin{eqnarray*}
		\frac{\partial  \bar\psi_\ep}{\partial t} &=& 
		2N \bar \psi_\ep \sum_{k=1}^5 [ A_k - \langle A_k \rangle  ] \frac{\partial\alpha^{\ast}_k}{\partial t} 
			+ \frac{\partial \delta}{\partial t}\,.
	\end{eqnarray*}
Next, we express the right-hand side of \eqref{PDE2} by artificially adding an expression $J_\ep[\bar \psi_\ep,p|\boldsymbol \alpha^\ast]$. This term represents dynamics where the forces, applied to the system, equal the effective forces. Due to stationarity the term equals zero. The reason for its inclusion is a cancellation of the diffusive part of the equation, that does not depend on $\boldsymbol \alpha$. This leads to a simple outcome where each term contains the difference between the true and the effective force
	\begin{eqnarray*}
		-\frac{\partial}{\partial p} J_\ep[\bar \psi_\ep,p | \boldsymbol \alpha] &=& 
		-\frac{\partial}{\partial p} J_\ep[\bar \psi_\ep,p | \boldsymbol \alpha]  + 
		\frac{\partial}{\partial p} J_\ep[\bar \psi_\ep,p | \boldsymbol \alpha^\ast] \\
		&=& -\frac{\partial}{\partial p} \left[ \frac{\sigma_\ep (p)}{2} 
		\sum_{k=1}^5 (\alpha_k - \alpha^\ast_k) \frac{\partial A_k}{\partial p} \bar \psi_\ep \right]
	\end{eqnarray*}
Equation \eqref{PDE2} thus becomes
 	\begin{eqnarray}\label{eq5by5}
		&&2N \bar \psi_\ep \sum_{k=1}^5 [ A_k - \langle A_k \rangle  ] \frac{\partial \alpha^\ast_k}{\partial t}  
		+ \frac{\partial \delta}{\partial t} = 
		\nonumber\\
		&&-\frac{\partial}{\partial p} \left[ \frac{\sigma_\ep (p)}{2} 
		\sum_{k=1}^5 (\alpha_k - \alpha^\ast_k) \frac{\partial A_k}{\partial p} \bar \psi_\ep \right] -\frac{\partial J_\ep [\delta,p]}{\partial p}
	\end{eqnarray}
where the function $\sigma_\ep(p) = pq$ in the bulk and $\ep q$ and $p \ep$ when $p<\ep$ and $p>1-\ep$, respectively.

\vskip 0.3cm
\noindent (c) We then multiply the equation  \eqref{eq5by5} by $A_j$ and average through allele frequency distribution. { The goal is to find  effective forces $\boldsymbol \alpha^\ast$ such that the error terms vanish, i.e. the projection of the full dynamics to the space of macroscopic quantities $\mathbf A$ is closed. Since there are $k$ constraints and the same number of forces, such $\boldsymbol \alpha$ in principle exist and are unique. The most crucial implication is that the approximation of macroscopic quantities is forced to coincide with the exact values $\mathbf A$ while the quasi-stationarity is valid. The equation becomes}
	\begin{eqnarray}
		&&2N  \sum_{k=1}^5 \frac{\partial \alpha^\ast_k}{\partial t} \int_0^1  [ A_k A_j - \langle A_k \rangle A_j  ] \bar \psi_\ep dp   = 
		\nonumber\\
		&&- \sum_{k=1}^5 \int_0^1 A_j \frac{\partial}{\partial p} \left[ \frac{\sigma_\ep (p)}{2} 
		 (\alpha_k - \alpha^\ast_k) \frac{\partial A_k}{\partial p} \bar \psi_\ep \right] dp
	\end{eqnarray}
Next, we use  integration by parts in the right-hand side of the equation. This introduces boundary terms, { i.e. terms evaluated at $p=0,\ep,1-\ep,1$ coming from the integration by parts}. We neglected these terms by assuming a rapid convergence to the stationary distribution and instantaneous adjustment of the { fluxes at the boundaries of the subdomains to their} stationary values. This stationarity assumption is adopted to avoid dependence of the approximation on the microscopic details of the distribution. This leads to to a relationship between the moment dynamics and dynamics of Lagrange multipliers in the full model
	\begin{equation}
		{2N} {\mathbf C}^\text{full}_{\ep} \frac{\partial \boldsymbol \alpha^{\ast}}{\partial t}  
		=  {\mathbf B}^\text{full}_\ep( \boldsymbol \alpha-\boldsymbol \alpha^\ast) \label{approx}
	\end{equation} 
where  $\boldsymbol \alpha = \boldsymbol \alpha^\text{full}$ and symmetric matrices ${\mathbf B}^\text{full}_\ep = \left \langle \frac{\sigma_\ep (p)}{2} \frac{\partial A_j}{\partial p} \frac{\partial A_k}{\partial p} \right \rangle$ and ${\mathbf C}^\text{full}_\ep = \text{Cov}(A_j,A_k)$  are obtained by averaging against the stationary $\ep$-dependent distribution 
	\begin{align}
		{\mathbf B}^\text{full}_\ep &= \left(
		\begin{array}{ccccc}
			\multicolumn{3}{c}{\multirow{3}{*}{\Huge $ {\mathbf B}_\ep$}}  & 0 & 0 \\ 
			&&&0&0\\
			&&&0&0\\
			0&0&0&0&0\\
			0&0&0&0&0
		\end{array}
		\right)\,, \\
		{\mathbf C}^\text{full}_\ep &= \left(
		\begin{array}{ccccc}
			\multicolumn{3}{c}{\multirow{3}{*}{\Huge ${\mathbf C}_\ep$}}  & \text{Cov}(\bar z,\chi_{[0,\ep]}) 
			& \text{Cov}(\bar z,\chi_{[1-\ep,1]}) \\ 
			&&& \text{Cov}(U_\ep,\chi_{[0,\ep]}) & \text{Cov}(U_\ep,\chi_{[1-\ep,1]})\\
			&&& \text{Cov}(V_\ep,\chi_{[0,\ep]}) & \text{Cov}(V_\ep,\chi_{[1-\ep,1]})\\
			\ast&\ast&\ast&Q(1-Q)&-PQ\\
			\ast&\ast&\ast&-PQ&P(1-P)
		\end{array}
		\right) \label{BC5}
	\end{align}
with ${ \mathbf B}_\ep$ and ${\mathbf C}_\ep$ are defined as ${\mathbf B}$ and ${\mathbf C}$ but in the context of the truncated stationary distribution and constraints, and with $\ast$ denoting the symmetric terms. 
The boundary masses entering into matrix ${\mathbf C}_\ep$ are defined as 
	\begin{align}
		Q &= \int_0^\ep \psi_\ep (p) dp \,, \qquad P = \int_{1-\ep}^1 \psi_\ep (p) dp\,.
	\end{align}
The matrix ${\mathbf B}_\ep^\text{full}$ is padded by zeros because $\partial_p \chi_{[0,\ep]}$ and $\partial_p \chi_{[1-\ep,1]}$ vanish in the interior of the subdomains. By neglecting the boundary terms { coming from the integration by parts} in the above calculation we effectively limit the transfer of the mass between the bulk and the boundaries so that the boundary masses agree with the continuous stationary distribution.

Note that the stationary allele frequency distribution $\bar \psi_\ep (\mathbf p)$ depends on the allele frequencies but also on the effective forces $\boldsymbol \alpha^\ast$. Therefore the matrices $\mathbf B_\ep$, $\mathbf C_\ep$, as well as matrices $\mathbf B^\text{full}_\ep$, $\mathbf C^\text{full}_\ep$, obtained by averaging over the microscopic states, are still functions of effective forces as in the continuous DynMaxEnt method \ref{apBC}. However, because the averages require integration through the interior domain  $[\ep,1-\ep]$ and separately through the boundaries $[0,\ep]$ and $[1-\ep,1]$, the integrals can no longer be written using the special functions and have to be evaluated numerically.

So far we have derived an extended dynamical system for the Lagrange multipliers $\boldsymbol \alpha^\ast$. One of the key questions is whether the dynamics converges to the target state $\boldsymbol \alpha$. Since the system has a form 
	\begin{equation}
		\frac{\partial \boldsymbol \alpha^{\ast}}{\partial t}  
		=  -{ {\mathbf M}^\ast_\ep} ( \boldsymbol \alpha^\ast-\boldsymbol \alpha) 
	\end{equation} 
where ${ {\mathbf M}^\ast_\ep}  = \frac{1}{2N} ({\mathbf C}^\text{full}_{\ep})^{-1} {\mathbf B}^\text{full}_\ep$ (note that the standard form of the ODE has a negative sign in front of ${ {\mathbf M}^\ast_\ep} $), convergence to the steady-state is captured by the sign of the eigenvalues of ${ {\mathbf M}^\ast_\ep} $. When they are positive (i.e. $-{ {\mathbf M}^\ast_\ep} $ has negative eigenvalues), the target state is asymptotically stable. It is easy to see that both  ${\mathbf B}^\text{full}_\ep$ and ${\mathbf C}^\text{full}_\ep$ are symmetric and positive semidefinite (${\mathbf C}^\text{full}_{\ep}$ is a covariance matrix and $\mathbf{u}^t {\mathbf B}^\text{full}_{\ep} \mathbf u>0$ can be written as a square) and thus the matrix ${ {\mathbf M}^\ast_\ep} $ has non-negative eigenvalues.
Therefore the dynamics of Lagrange multipliers should converge to the fixed point. This fixed point is precisely the target point $\boldsymbol \alpha$ unless some of the eigenvalues converge to zero -- then the dynamics may get trapped at a different point of the phase space  and are unable to continue all the way to the target state. 

Moreover, because the dynamics of Lagrange multipliers is five-dimensional while the dynamics of observables, driven by the ${\mathbf B}$-matrix has only three degrees of freedom (since the entries of ${\mathbf B}^\text{full}_\ep$, corresponding to $A_4$, $A_5$ are zero), the method is in principle underdetermined. This essentially means that Lagrange multipliers $\mu$ and $\kappa$ and $\nu$ and ${ \rho}$ may follow strange dynamics that together nevertheless lead to well-approximated observables. This is why in the next step we need to employ the constraints on $\kappa$ and $\rho$ by slaving their dynamics with dynamics of mutation rates
	\begin{align}\label{kapparho}
		\kappa &= -\frac{1}{2N} \log (4N\!\mu) \,, \qquad \rho = -\frac{1}{2N} \log (4N\!\nu)
	\end{align}
Imposing these constraints leads to a reduced dynamics with only three degrees of freedom, 
$ {\bf A} = ( \bar z, U_\ep , V_{\ep})$ and $ {\boldsymbol  \alpha} = \left(\boldsymbol \beta ,\!\mu,\!\nu \right)$.

The full dynamics can be reduced by disregarding the entries of ${\mathbf B}^\text{full}_\ep$ that are identically equal to zero and considering only the submatrix ${\mathbf B}_\ep$. Intuitively, one expects that the reduced dynamics will be identical to the dynamics \eqref{QSDA} governed by $3\times 3$ matrices ${\mathbf B}_\ep$ and ${\mathbf C}_\ep = \text{Cov}(A_j,A_k)$.  However, we show that the ${\mathbf C}$-matrix involves additional terms.
To do that, we start by taking the first 3 equations from \eqref{eq5by5}, leaving out the equations for the dynamics of $\rho$ and $\kappa$ { and by substituting $\rho$ and $\kappa$ by the expressions of $\mu$ and $\nu$ from \eqref{kapparho}}
	\begin{align*}
		2N \sum_{k=1}^3 \frac{d\alpha^\ast_k}{dt}  \psi [A_k-\langle A_k \rangle]] 
		- \sum_{k=2}^3 \frac{1}{\alpha^\ast_k}\frac{d\alpha^\ast_k}{dt}  \psi [A_{k+2}-\langle A_{k+2} \rangle]]  + \partial_t \delta =\\
 		-\partial_p \left[ \frac{\sigma(p)}{2} \sum_{k=1}^3 \left((\alpha-\alpha^\ast)\frac{\partial A_k}{\partial p} \right) \right] -
		 \partial_pJ[\delta,p]\,.
	\end{align*}
We are left with a system of three ODE's that involve terms from the dynamics of $\kappa$ and $\rho$ { (the second sum on the left)}. { We then substitute values of $\kappa$ and $\rho$ from \eqref{kapparho} and obtain the correct ${\mathbf C}$-matrix  }
	\begin{align}
		{\mathbf C}_\ep&= 
		\begin{pmatrix}
			\text{Var}(\bar z) &  \text{Cov}(\bar z,U_\ep)
			& \text{Cov}(\bar z,V_\ep)  \\
			\text{Cov}(\bar z,U_\ep)&  \text{Var}(U_\ep)
			& \text{Cov}(U_\ep,V_\ep) \\
			\text{Cov}(\bar z,V_\ep) &{ \text{Cov}(U_\ep,V_\ep)}
			&\text{Var}(V_\ep)
		\end{pmatrix}  \\
		&-\begin{pmatrix}
			0 &  \frac{1}{2N\!\mu} \text{Cov}(\bar z,\chi_{[0,\ep)}) 
			& \frac{1}{2N\!\nu} \text{Cov}(\bar z,\chi_{[1-\ep,1)})  \\
			0& \frac{1}{2N\!\mu} \text{Cov}(U_\ep,\chi_{[0,\ep)}) 
			& \frac{1}{2N\!\mu} \text{Cov}(U_\ep,\chi_{[1-\ep,1)}) \\
			0 &\frac{1}{2N\!\mu} \text{Cov}(V_\ep,\chi_{[0,\ep)})  
			&\frac{1}{2N\!\nu} \text{Cov}(V_\ep,\chi_{[1-\ep,1)})
		\end{pmatrix}\,, \nonumber
	\end{align}
resulting in a reduced dynamics $\frac{\partial \boldsymbol \alpha^\ast}{\partial t}= \frac{1}{2N}  ( {\mathbf  C}_\ep)^{-1} {\mathbf B}_\ep( \boldsymbol \alpha-\boldsymbol \alpha^\ast)$. Note that despite the original intuition, the added terms in the second and third column must be present to reflect the dynamical character of the prefactors in the stationary distribution.
The case with dominance is treated in an analogous way and results in matrices $\mathbf B_\ep$ and $\mathbf C_\ep$ of dimension 4 by 4.

An important question is whether dynamics described in the above system converge to the state ${\boldsymbol \alpha}$. 
This is true for the dynamics of observables, driven by matrix ${\mathbf B}_\ep$,  no matter  the initial and target state, because of the symmetry and  positive definiteness. This means that the approximation can in principle work, but it remains to be seen if  the dynamics of Lagrange multipliers also have good convergence properties. The matrix $ {\mathbf C}_\ep$ is no longer symmetric and thus $ ({\mathbf C}_\ep)^{-1} B_\ep$ is not necessarily positive definite. However, all our simulations suggest that the positive semidefiniteness holds and the only way how the approximation may fail to converge to the fixed point $\boldsymbol \alpha$ is when some of the eigenvalues go to zero. In such case the trajectory gets trapped at a different state. We have observed this behaviour for $4N\!\mu \ll 1$ where one of the mutation rates approaches zero.

\section{Dependence on the truncation parameter $N\!\ep$} \label{apep}

Evolution of quantitative traits depends on the following non-dimensional parameters: $N\!\beta$, $N\!\mu$ and $N\!\nu$. The general DynMaxEnt approximation, as well as the piecewise-defined diffusion \eqref{PDE2} depends on an additional nondimensional parameter $N\!\ep$ that influences the quality of the general DynMaxEnt approximation. As this parameter gets very small, the method approaches the continuous method without boundary layers. Since in the limiting case matrix ${\mathbf B}_\ep \rightarrow {\mathbf B}$ contains diverging components, the error of the general approximation should increase as $N\!\ep$ gets smaller. Similarly, the error will be large when { $N\!\ep \gg 1$}. This is because the relationship \eqref{SCdist4}, which assumed $N\!\ep\ll 1$, no longer matches the boundary masses of the approximated stationary distribution with the correct stationary distribution. Moreover, the boundary is too wide for $N\!\ep \gg 1$ to disentangle effects of selection and mutation. 

\begin{figure}[h]
		\centering
		\includegraphics[width=0.98\textwidth,keepaspectratio]{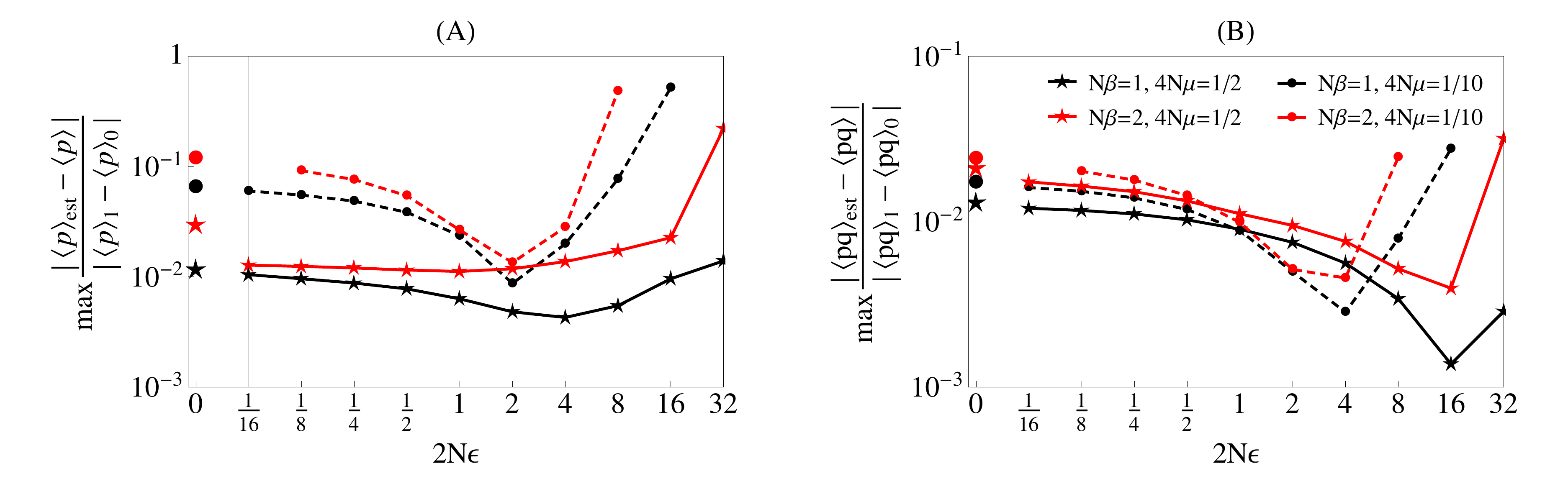}
		\caption{ 
			Effect of $N\ep$ on performance of the general DynMaxEnt  method. 
			The response to a fast change in selection $N\!\beta = -1$ to 1 (or -2 to 2) has been compared 
			with the approximated response of the system while keeping mutation fixed $4N\!\mu = 4N\!\nu = 1/2$ (or 1/10) 
			and no overdominance $N\!h=0$ and $2N=1000$. Each panel shows all four combinations of parameter regimes, 
			distinguished by color ($N\!\beta$) and filling ($4N\!\mu$). 
			We used a method that constrains expectations of three observables, 
			$\langle \bar z \rangle$, $\langle U \rangle$ and $\langle V \rangle$. 
			Panels also contain the error of the continuous DynMaxEnt method, plotted at $N\ep = 0$, that was obtained by fixing
			effective mutation $\mu^\ast = \mu$, $\nu^\ast=\nu$.
			(A) Relative error of the trait mean; 
			(B) Relative error of the mean heterozygosity.
		\label{ex_eps1}}
	\end{figure}

Simulations in Figure~\ref{ex_eps1} applied a sudden change in selection while the small mutation rates remained unperturbed. The figure shows that there exists an optimal value of the threshold $N\!\ep$ that leads to the best approximation. This threshold seems to  depend on the mutation rates $N\!\mu, N\!\nu$ but also weakly on the selection rate $N\!\beta$, which is consistent with a representation of $p<N\!\ep$ as a mutation-dominated regime. The optimal threshold differs when considering error in the trait mean and in the mean heterozygosity.

\begin{figure}[h]
		\centering
		\includegraphics[width=0.98\textwidth,keepaspectratio]{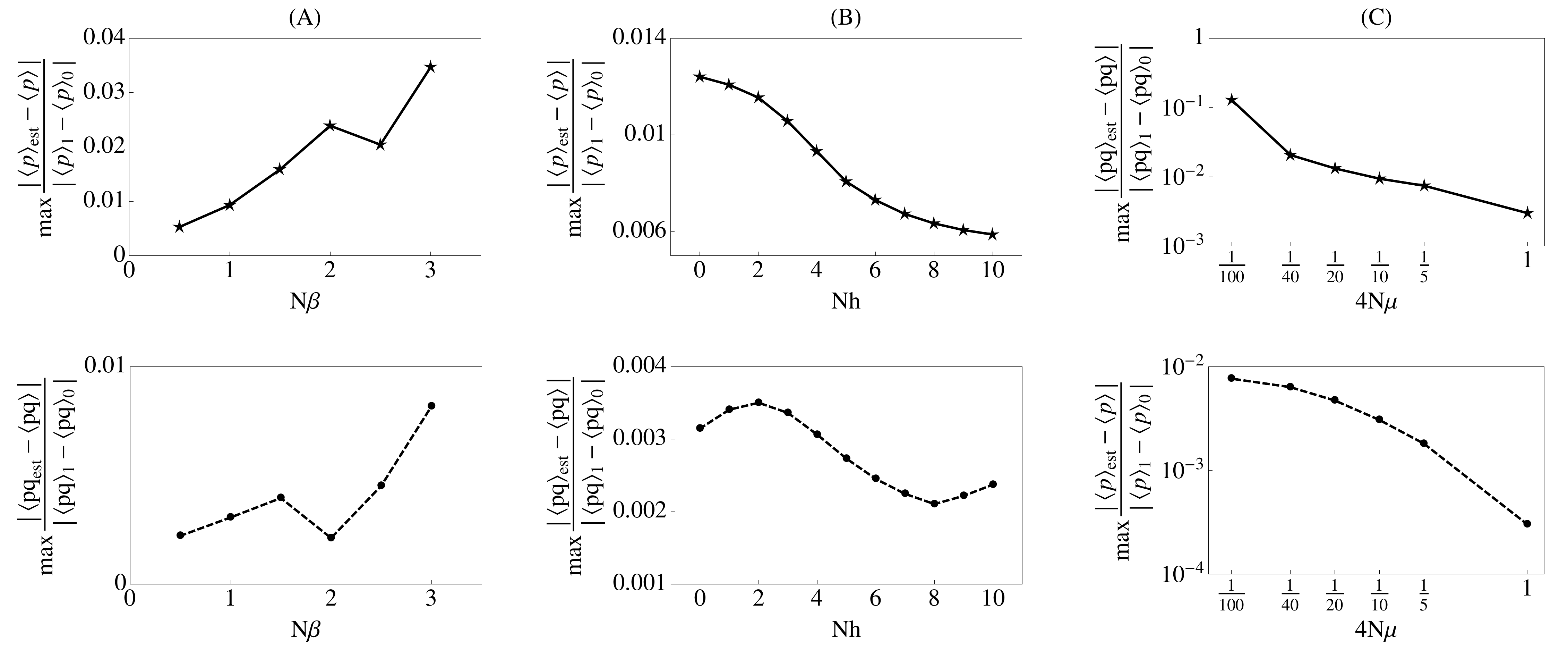}
		\caption{ 
			Effect of $N\boldsymbol\alpha$ on performance of the general DynMaxEnt  method. 
			The reference simulation undergoes a fast change in selection $N\!\beta: -1\rightarrow 1$, 
			heterozygosity $N\!h: 4\rightarrow 0$ and mutation $4N\!\mu, 4N\!\nu: 0.5 \rightarrow 0.1$  and $2N=1000$.
			The three panels show dependence of the simulation error when the reference case parameters are 
			changed one at a time.
			(A) Dependence on the strength of directional selection, changing from $-N\!\beta$ to $N\!\beta$
			(plotted on a linear scale).
			(B) Dependence on the strength of selection for the heterozygous form, changing from $Nh$ to 0 
			(plotted on a linear scale).
			(C) Dependence on the magnitude of mutation, changing from 0.5 to $4N\!\mu = 4N\!\nu$ 
			(plotted on a logarithmic scale). 
		\label{ex_eps2}}
	\end{figure}

It is useful to know how the general DynMaxEnt method performs for realistic examples given a sensible fixed choice of $N\ep$. 
We set $2N\ep = 1$, such that $\ep$ corresponds to a frequency of one in the total of $2N$ individuals. We took the reference simulation in Example 2 and varied strengths of the parameters $N\!\beta$, $Nh$, $4N\!\mu = 4N\!\nu$ one at a time. Figure~\ref{ex_eps2} shows that the relative error in a trait mean is for most parameter settings within 1 percent. The worst case scenario is when selection gradient is strong and mutation  drops rapidly to a small value. 
This is expected, since strong and rapid change in selection implies a large deviation from adiabatic regime, thus the quasi-stationarity assumption may fail. Additionally, small mutation leads to a separation of timescales in the model that may be too complex to capture by a  simple approximation. Surprisingly, stronger selection on heterozygosity does not decrease the accuracy of the method, which is counterintuitive because large $Nh$ leads to more complex allele frequency distributions.

\section{Comparison of different methods} \label{apComp}

We compare three available methods: the discrete approximation (see Appendix~\ref{apD}), that captures response to a change in selection in the limit $4N\!\mu, 4N\!\nu \rightarrow 0$; the continuous DynMaxEnt approximation (see \eqref{QSDA}), valid for $4N\!\mu, 4N\!\nu >1$ with a possibility to treat also small but static mutation; and the general DynMaxEnt approximation (see \eqref{eqaep}), that extends the validity to dynamic mutation of an arbitrary magnitude. Note that dynamic Lagrange multipliers may also represent changing population size, not just the evolutionary forces.

	\begin{figure}[h]
		\centering
		\includegraphics[width=0.32\textwidth,keepaspectratio]{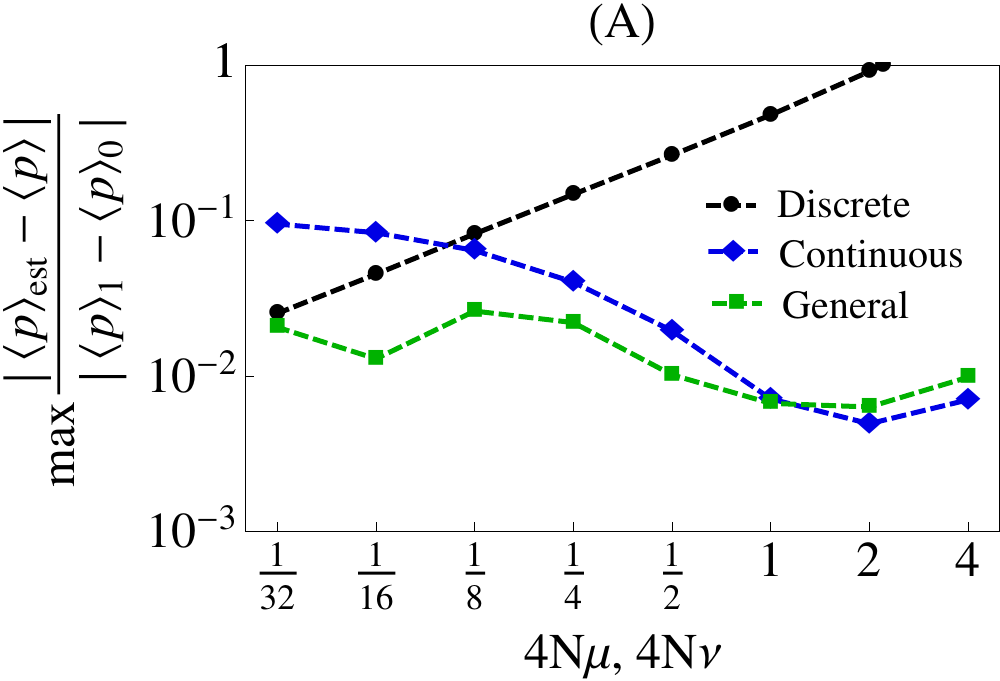}
		\includegraphics[width=0.32\textwidth,keepaspectratio]{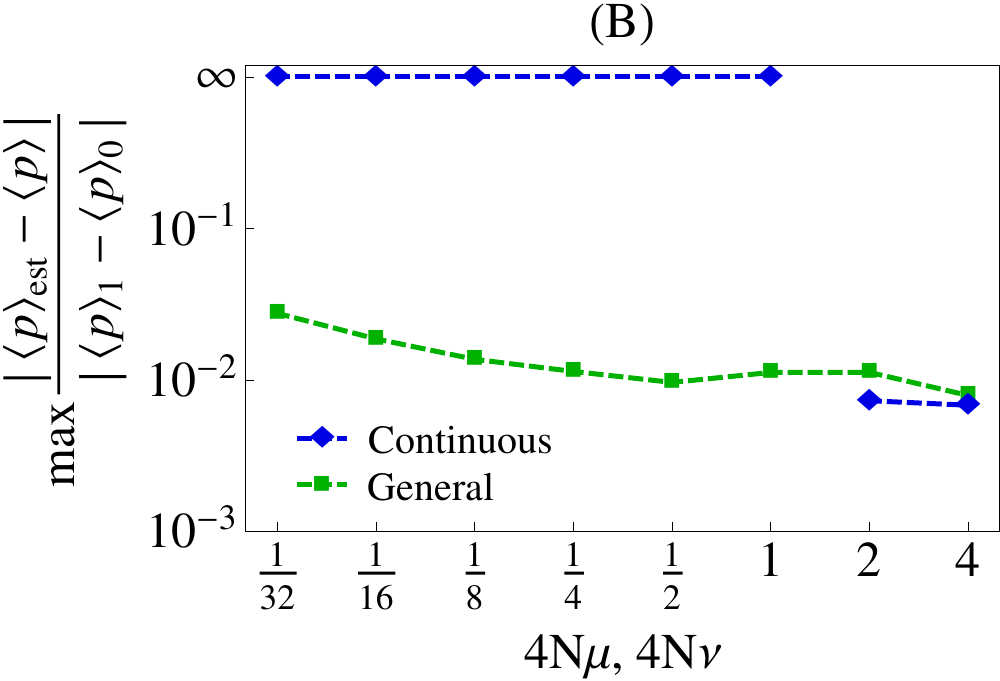}
		\caption{\label{figComp} 
		Comparison between discrete, continuous and general approximations. We show performance of the methods in 
		two scenarios: 
		(A) Rapid change in selection $N\!\beta = -1$ to 1 with other parameters unperturbed $Nh = 0$, 
		$N\!\mu = N\!\nu \in \{ 2^{-5}, \dots, 2^2 \}$. 
		(B) Rapid change in selection $N\!\beta = -1$ to 1 and mutation $4N\!\mu = 4N\!\nu = 1.1$ to $ \{ 2^{-5}, \dots, 2^2 \}$,
		combined with a change in heterozygosity $Nh = 4$ to 0. The population size is $2N=1000$.}
	\end{figure}

Figure~\ref{figComp} shows two scenarios: perturbation of the system through a change of selection (panel A), and perturbation through changes in all evolutionary parameters (B). In the case (A) the continuous DynMaxEnt method can be applied even in the case of small mutation $4N\!\mu<1$ by enforcing fixed mutation rate in time. But even then, the continuous method performs worse than the general method, which shows a relative error of the order of few percent. In comparison, the discrete approximation gives an accurate estimate when mutation drops below $4N\!\mu<0.1$ but due to its simplicity it does not capture unequal and dynamic mutation rates. 

The power of the general DynMaxEnt is demonstrated in  Figure~\ref{figComp} (B) for dynamic mutation. This situation cannot be captured by the simple discrete approximation and the continuous DynMaxEnt, not valid for $4N\!\mu<1$, shows divergence that is an inherent property of the matrix $\mathbf B$, approximating dynamics of observables. To show performance of the continuous method we included $\infty$ in the axis denoting accuracy of the approximation.

\section{Irreversibility of the adaptation dynamics} \label{apI}

\begin{figure}[h]
		\centering
		\includegraphics[width=0.98\textwidth,keepaspectratio]{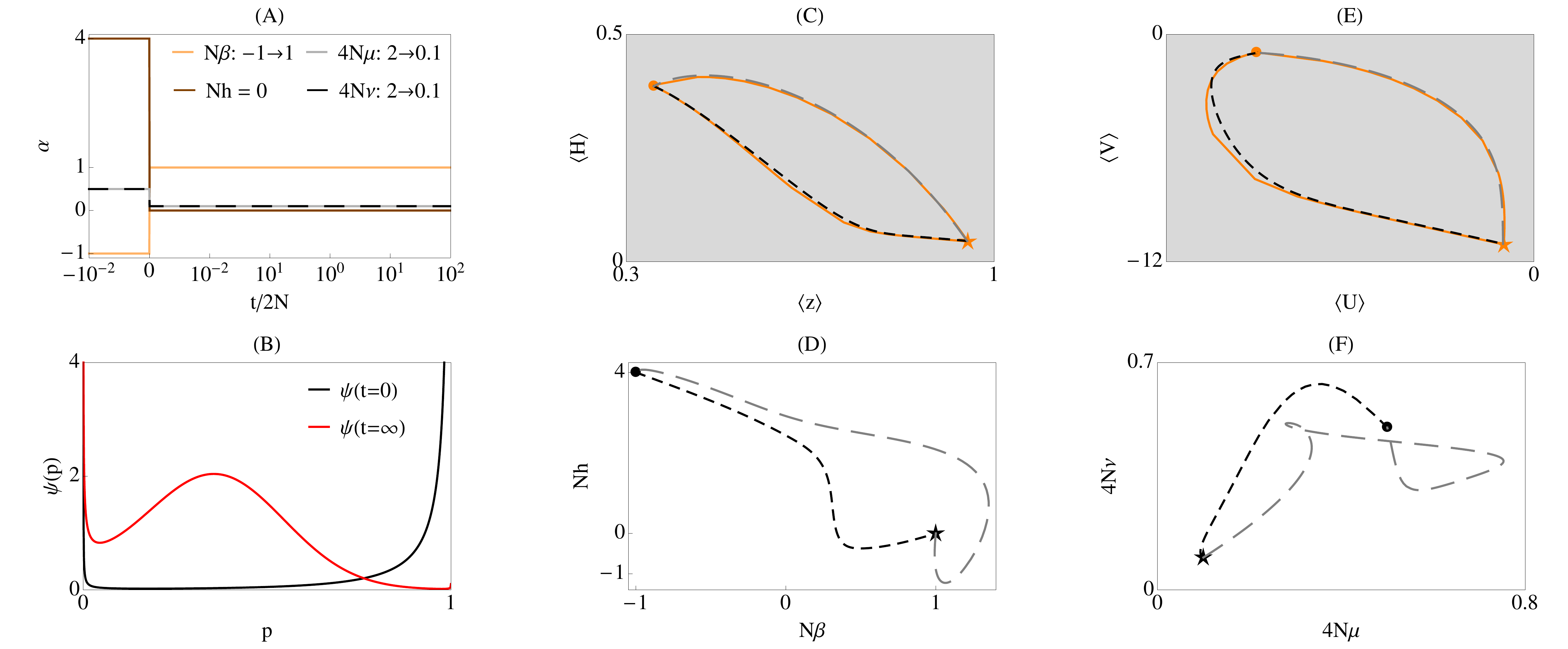}
		\caption{ 
			Irreversibility of the dynamics for Example 2. 
			The dynamics in Figure~\ref{fig_ex2f1} have been complemented by dynamics with the initial state 
			and the final state reversed.  
			(A) Values of evolutionary forces for the reversed simulation.
			(B) Initial and final allele frequency distribution for the reversed simulation.
			(C, E) The paths  between the initial/final conditions (circle/star) displayed in the space of observables, 
			together with the exact dynamics, showed in orange. 
			We displayed the situation in two out of six possible pairs of coordinates. 
			(D, F) The dynamics, visualized in the space of corresponding Lagrange multipliers. 
			All figures show the forward trajectory (Example 2) in gray-dashed curve and dynamics with 
			reversed initial and final state in black-dashed curve.}
		\label{ex4fig1}
	\end{figure}

Figure~\ref{ex4fig1} { compares} the dynamics of the observables and Lagrange multipliers for Example 2, where the initially tri-modal distribution loses polymorphism due to a decrease in the selection for heterozygous individuals and due to a simultaneous decrease in the mutation rate, and a reversed dynamics with the initial and final conditions switched. The dynamics { are} irreversible since switching the initial and final conditions leads to very different paths in the space of observables as well as in the space of the effective forces. The dynamics in the space of observables { show} a good agreement with the exact solution of the problem. The paths from the initial to the final state are relatively straight. On the contrary, the dynamics of Lagrange multipliers shows a complicated response of the system, that exhibits a separation of timescales and overshooting of the  equilibrium levels; for instance, the effective selection coefficient grows to a level $N\!\beta \approx 1.5$ before decreasing back to $N\!\beta = 1$. 

\section{Equivalence between single locus approximation and multiple loci with equal effects} \label{apE}

A trait typically depends on many loci with different effects. If these effects are called $\gamma_i$, the simplest case is an additive trait $\bar z = \sum_i \gamma_i (p_i-q_i)$. If linkage equilibrium is assumed the constraints can be written in terms of allele frequencies at the loci where each constraint is additive through loci ($A_i = \sum_k^n A^{(k)}_i$) with symmetric expressions in terms of $p_i$. While the selection-related constraints depend on the distribution of effects $\gamma$ linearly, the mutation-related constraints do not depend on it. If the effects of the ${ L}$ loci on an additive trait are the same and $\gamma_i=1$ then
	\begin{align*}
		{ \mathbf B}_{ L} &= \left\langle  \sum_{k=1}^{ L} \frac{\partial \mathbf A}{\partial p_k} \frac{p_k q_k}{2} 
		\frac{\partial \mathbf A^{T}}{\partial p_k} \right\rangle
		=  \sum_{k=1}^{ L}  \left\langle \frac{\partial \mathbf A}{\partial p_1} \frac{p_1 q_1}{2} 
		\frac{\partial \mathbf A^{T}}{\partial p_1} \right\rangle
		= { L} { \mathbf B}_1
	\end{align*}
where ${\mathbf B}_{ L}$ is the matrix of genetic covariances for ${ L}$ loci. This is true even in the approximation for small dynamic mutation since the boundary terms are also additive accross loci. A similar relationship holds for the covariance of fluctuations ${\mathbf C}_{L} = {L}{\mathbf C}_1$. Therefore, the matrix  defining the maxent dynamics of Lagrange multipliers in case of ${ L}$ loci with equal effects is identical to dynamics of Lagrange multipliers in case of a single locus, i.e. ${\mathbf C}_{L}^{-1}{ \mathbf B}_{L}={\mathbf C}_1^{-1}{ \mathbf B}_1$. As a result, the trait mean $\bar z_n = \sum_{k=1}^{L} (p_i-q_i) = {L}(p-q)=n\bar z_1$, where $p$ represents the allele frequency in the single locus simulation with other parameters unchanged. On the other hand, the calculation of a trait mean for ${L}$ independent loci of equal effects $\gamma_i=1$ using the continuous model gives
	\begin{align}
		\frac{d\bar z_{L}}{dt} &= 2\sum_{k=1}^{L} \frac{dp_i}{dt} = 
		{L}\frac{d\bar z_1}{dt}
	\end{align}
showing that the relationship $\bar z_{L} = {L} \bar z_1$ follows also from the model description.

\section{Additional results for multiple loci with unequal effects} \label{ap_multi}
\begin{figure}[h]
		\centering
		\includegraphics[width=0.98\textwidth,keepaspectratio]{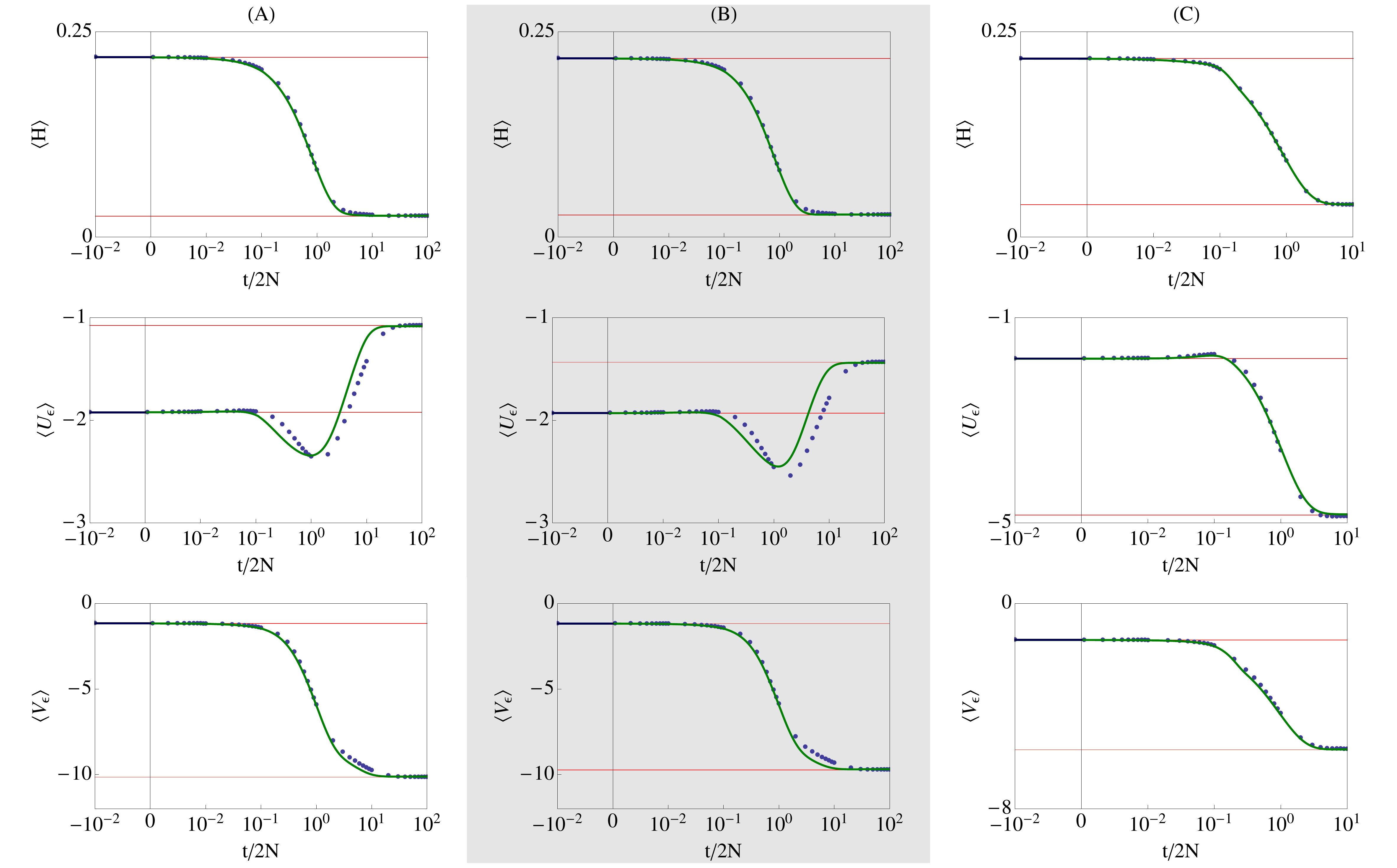}
		\caption{    {Example 3,} dynamics of the quantitative traits for 100 loci of different effects. 
			The effects are randomly drawn from: (A) uniform distribution in $[0,2]$; (B) exponential distribution with mean 1; and 
			(C) deterministic effects where 95 of the loci have effects 0.01 while the remaining five loci have effects 19.81. 
			The true forces $\boldsymbol \alpha$ change at time $t=0$ and draw the system out of equilibrium. 
			The response of the observed quantities, $\langle pq \rangle$, $\langle 2\log p \rangle$ and 
			$\langle 2 \log q \rangle$, is shown in blue dots while 
			the approximation is shown in green. The quality of the approximation for the trait mean $\langle \bar z \rangle$ 
			is shown in Figure~\ref{fig_ex3f3}.
		\label{fig_ex3f4} }
	\end{figure}

{ Figure~\ref{fig_ex3f4} shows details on the quality of the general DynMaxEnt approximation in Example 3, where a linear trait, depending on 100 loci with different effects, evolves due to a combination of a directional selection and selection on heterozygosity, random drift and mutation. The changes in  evolutionary forces are the same as in Example 2.}

\end{document}